\documentclass[preprint,11pt]{elsarticle}

\usepackage{lineno,hyperref}
\modulolinenumbers[5]
\usepackage[utf8]{inputenc}
\usepackage{eurosym}
\usepackage{enumitem}
\setlist[description]{leftmargin=\parindent,labelindent=\parindent}
\usepackage{amsmath}
\usepackage{amsfonts}
\usepackage{amssymb}
\usepackage{amsthm}
\usepackage{breqn}
\usepackage{float}
\usepackage{multirow}
\usepackage{graphicx}
\usepackage[table,xcdraw]{xcolor}
\usepackage{tabularx}
\usepackage[para]{threeparttable}
\usepackage{adjustbox}
\usepackage{makecell}
\usepackage{rotating}
\usepackage{lscape}
\usepackage{cleveref}
\usepackage{pgfplots}
\usepgfplotslibrary{statistics}
\usepackage{soul} 
\usepackage{arydshln}
\usepackage{tikz}
\usetikzlibrary{shapes,arrows}
\usepackage{standalone}
\usepackage{import}
\usepackage{caption}
\usepackage{subcaption}
\usepackage{graphicx}
\pgfplotsset{compat=1.17}

\usepackage[a4paper, total={16cm, 22cm}]{geometry}

\crefname{subsection}{subsection}{subsections}

\makeatletter
\def\ps@pprintTitle{%
  \let\@oddhead\@empty
  \let\@evenhead\@empty
  \def\@oddfoot{\reset@font\hfil\thepage\hfil}
  \let\@evenfoot\@oddfoot
}
\makeatother

\biboptions{authoryear}

\begin{document} 

\begin{frontmatter}

\title{A multiple criteria approach for ship risk classification: \\ An alternative to the Paris MoU Ship Risk Profile
}

\author[myprimaryaddress]{Duarte Caldeira {\sc{Dinis}}\corref{mycorrespondingauthor}}
\cortext[mycorrespondingauthor]{Corresponding author}
\ead{duarte.dinis@tecnico.ulisboa.pt}

\author[myprimaryaddress]{José Rui {\sc{Figueira}}}
\ead{figueira@tecnico.ulisboa.pt}

\author[mysecondaryaddress]{\\Ângelo Palos {\sc{Teixeira}}}
\ead{teixeira@centec.tecnico.ulisboa.pt}

\address[myprimaryaddress]{CEG-IST, Instituto Superior Técnico, Universidade de Lisboa, \\ Av. Rovisco Pais 1, 1049-001, Lisboa, Portugal}

\address[mysecondaryaddress]{CENTEC, Instituto Superior Técnico, Universidade de Lisboa, \\ Av. Rovisco Pais 1, 1049-001, Lisboa, Portugal}

\begin{abstract}
\noindent The Paris Memorandum of Understanding on Port State Control (Paris MoU) is responsible for controlling substandard shipping in European waters and, consequently, increasing the standards of safety, pollution prevention, and onboard living and working conditions. Since 2011, the Memorandum adopted a system of points, named the ``Ship Risk Profile'' (SRP), under which each ship is assigned a risk profile according to its score on a set of criteria. Being a multiple criteria decision aiding (MCDA) tool at its core, comprising criteria, weights, and risk categories, limited research has been performed on the SRP from an MCDA perspective. The purpose of this paper is to propose an MCDA approach for ship risk classification through the Deck of Cards Method (DCM). The DCM is particularly suitable within this context as it allows, intuitively for the decision-maker, to model preference among different criteria and among different levels on criteria scales. First, a framework is built, based on the criteria established in the SRP. Second, the DCM is used to build the MCDA model, including the definition of the criteria value functions and criteria weights. Finally, the proposed MCDA model is applied to a dataset of ships and the results are discussed. Robust results have been obtained with the proposed approach, making it a potential alternative to the current SRP.  
\end{abstract}


\end{frontmatter}


\section{Introduction}
\label{section:introduction}
\noindent The Paris Memorandum of Understanding on Port State Control (Paris MoU) of 1982 is considered to be the first regional agreement on Port State Control (PSC) \citep{Bang2012}, instituting a coordinated system of inspections by the maritime authorities of its member States \citep{Graziano2017}. The success of the inspection system resides on the sharing of information about ships between authorities, avoiding the unduly inspection of the same ship in different ports and enabling the identification of ``delinquents'' \citep{Hare1997}. In addition, it provides harmonised rules and standards for the selection of vessels to be inspected, and regarding the inspection and detention procedures \citep{Graziano2017,Graziano2018}. The 25\% inspection quota for each member State established in the Paris MoU of 1982 \citep{Lowe1982} was replaced in 2011 with the adoption of the New Inspection Regime (NIR) in the 32nd Amendment to the Paris MoU \citep{ParisMoU2012}. The NIR is based on a system of points, named the ``Ship Risk Profile'' (SRP), in which ships are attributed points according to a set of criteria. The obtained score determines the ``risk profile'' of a given ship, which, in turn, determines its priority for inspection, the interval between inspections, and the scope of the inspection. Notwithstanding ``overriding and unexpected factors'' that may trigger additional inspections \citep{ParisMoU2020}, the principle behind the NIR is that the lower the risk a ship poses, the longer the inspection interval and less detailed the inspection.

The SRP structure comprises a set of ``risk profiles'', a set of “criteria'' with associated ``weighting points'', and a set of ``parameters''. These elements have inspired research on other aspects related to maritime transport. \cite{Yang2018a} use some of the SRP parameters, collected from the Paris MoU online inspection database, to propose a model based on Bayesian networks (BNs) with the aim of helping port authorities in determining the optimal ship inspection policy. \cite{Yang2018b} use the SRP parameters in a model to analyse risk influencing factors in PSC inspections and to predict the probability of ship detention. Based on the work of \cite{Sage2005}, \cite{Dinis2020} propose a model that use the SRP parameters as risk variables for ship and maritime traffic risk assessment with BNs. Notwithstanding the importance of PSC in controlling substandard shipping and increasing the standards of safety, pollution prevention, living and working conditions, particularly since the implementation of the NIR \citep{Yang2020}, limited research has been devoted to analyse the SRP and its structural elements – risk profiles, criteria, weights, and parameters \citep{ParisMoU2020}.

Being a multiple criteria decision aiding (MCDA) tool at its core, limited research has been performed on the SRP from an MCDA perspective. MCDA deals with decision problems with a finite or infinite number of actions (i.e., alternatives, options, etc.), at least two criteria, and at least one decision-maker (DM). The problem involves the DM (or DMs) choosing, ranking, or sorting the actions according to their performances on a set of criteria \citep{Greco2016}. In the particular case of the SRP, ships, representing the actions, have to be assigned to the most appropriate risk category, according to their performance on the set of considered criteria. PSC authorities, representing the DMs, shall then decide the inspection interval to be assigned to each ship. In MCDA terms, this represents a sorting problem \citep{Greco2016}, to which a recent survey on methods used in solving it can be found in \cite{Alvarez2021}. Just like a formal MCDA sorting model, the SRP comprises a set of ordered categories (the ``ship risk profiles''), criteria and criteria scales (the ``parameters'' and ``criteria'' of the SRP, respectively), and model parameters, namely the ``relative importance'' of the criteria (the SRP ``weighting points''). Limited attention has been given in the literature, nonetheless, to the structural elements of the Paris MoU SRP. 

It is the purpose of this study to analyse these elements and to formalise them through an MCDA model. To achieve this purpose, the study employs a variant of the Deck of Cards Method (DCM) \citep{Simos1989,Figueira2002,Corrente2021}. The DCM allows the construction of ratio and interval scales, which can be used to determine the weights of criteria \citep{Figueira2002} and to accurately model the strength of preference between different levels in criteria scales \citep{Corrente2021}. In this case, it is employed to model the preferences regarding the criteria, i.e. the SRP ``parameters'', and the respective criteria scales, i.e. the SRP ``criteria''. Criteria weights are established through the former, while criteria value functions are established through the latter. From the performed literature review, several examples exist using the DCM as an auxiliary approach to define the criteria weights of outranking MCDA methods. On the contrary, this study employs the DCM to construct a complete MCDA model, from the definition of the criteria value functions, to the elicitation of the model's weights. To the best of the authors knowledge, this paper is the first proving the applicability of the DCM in performing such a task.

The DCM, as proposed by \cite{Simos1989}, later improved by \cite{Figueira2002}, and extended to other contexts by \cite{Corrente2021}, has been used in a multitude of sectors, namely for the determination of criteria weights in MCDA outranking methods, such as the \textsc{Electre} and \textsc{Promethee} methods \citep{Figueira2016}, as it was originally conceived. One of the sectors using the DCM for the definition of criteria weights is the energy sector. \cite{Haurant2011} used the DCM together with the \textsc{Electre IS} method for the selection of photovoltaic plant projects in the island of Corsica. \cite{Dawson2012} used the DCM to rank the suitability parameters for the installation of Concentrated Solar Thermal Power (CSP) plants in Australia and to allow the calculation of weight values. \cite{Neves2018} used the DCM with the \textsc{Electre III} in the development of a sustainable energy strategy at the municipal level. More recently, \cite{Riley2020} used the DCM with the \textsc{Electre III} to assess the social, economic, and environmental impacts of production development of methane gas hydrate in Alaska. Water supply is another sector in which the DCM has been used. \cite{Kodikara2010} used the DCM to elicit criteria weights with the \textsc{Promethee} MCDA outranking method \citep{Figueira2016} in an application to the water supply system of Melbourne, Australia. \cite{Mutikanga2011} combined the DCM with the \textsc{Promethee II} and applied the methods to water loss management in Kampala, Uganda. More recently, \cite{Pinto2017} developed a performance assessment model based on the \textsc{Electre Tri-nC} for water utilities in Portugal, using the DCM for criteria weighting. In banking, \cite{Angilella2015} and \cite{Doumpos2019} use the DCM to establish the criteria weights of MCDA models for credit rating, the former combining with the \textsc{Electre Tri}, and the latter with the \textsc{Electre Tri-nC}. \cite{DelVasto-Terrientes2015} propose an outranking MCDA method for hierarchically structured criteria, called \textsc{Electre-III-H}, and use the DCM to determine the criteria weights. A case study on the application of the method is provided regarding the construction of a priority ranking of tourism websites. Examples of other sectors applying the DCM in the definition of criteria weights for MCDA models include manufacturing \citep{Cavallaro2010}, organisational development \citep{Merad2013}, supplier selection in food supply chains \citep{Govindan2017}, soldier selection \citep{Costa2020}, but also maritime safety \citep{Silveira2021}. In the latter work, the authors combined the DCM with the \textsc{Electre Tri-nC} to assess the ship collision risk based on expert judgments, demonstrating the applicability of the DCM to the maritime sector.

Notwithstanding the diversity of examples in which the DCM has been employed, alternative methods have been proposed and applied in the literature for the determination of criteria weights in MCDA. \cite{Malekmohammadi2011} combined the \textsc{Electre Tri} with mathematical programming to infer model parameters, including criteria weights, from information provided by the DM to the model, such as examples on the assignment of alternatives to categories. \cite{Kaliszewski2016} proposed a new method to capture ``preference information'' from the DM, including criteria weights, based on the Simple Additive Weighting (SAW). \cite{Lolli2019} developed an approach to elicit criteria weights indirectly from partial rankings or from the selection of the most preferred alternative by the DM. The approach is combined with a \textsc{Promethee}-based ranking method and used to improve recommendations for mobile applications in smartphones. Not being exactly an alternative MCDA technique, \cite{Kadzinski2020} proposed a method to verify the consistency of elicited preferences, which extends the Segmenting Description (SD) approach. The method is put forward to identify inconsistencies between the DM's judgments and a preference model assumed \textit{a priori}.

The remainder of the paper is organised as follows. In Section \ref{section:methodology}, an introduction to the application of MCDA to the problem of risk classification in shipping is presented. In Section \ref{section:application}, the DCM is used in the construction of an MCDA model for ship risk classification. A framework is proposed based on the Paris Mou SRP criteria and the DCM is used to build criteria scales and to assign criteria weights. The proposed model is applied to a data sample of ships and the obtained results are discussed. In Section \ref{section:casestudy}, a case study is presented, in which the risk classification of a data set of ships is compared between that obtained with the proposed MCDA model and that defined through the Paris MoU SRP. Managerial insights on the application of the proposed approach are also provided. Finally, in Section \ref{section:conclusion}, conclusions are drawn on the performed work and future research is identified.

\section{An MCDA methodology for ship risk classification}
\label{section:methodology}
\noindent An important note has to be made at this point. Instead of the terms used in the Paris MoU SRP, a different terminology is employed in this paper, in line with the established definitions in MCDA literature. Instead of ``parameter'', this paper uses the term ``criterion''. Based on \cite{Roy1999}, a \textit{criterion} is a model ``for evaluating and comparing potential actions'', which occurs by assessing the performance of each action, in this case ships, on a ``preference scale''. The latter term or, equivalently, ``criterion scale'' is used in this paper instead of ``criterion'', as defined in the SRP. The term ``weight'' is used in this paper similarly to the term ``weighting'' of the SRP. However, in practice, the current SRP is based on a simple sum of points according to the performance of each ship on each criterion, in which the ``weighting points'' do not have the intrinsic characteristic of measuring the relative importance of criteria \citep{Figueira2002}. In MCDA terminology, criteria weights are considered as ``preference parameters'' of MCDA models \citep{Corrente2021}. 

As previously mentioned, this paper employs the DCM to assess the weights of criteria on ratio scales and to built the value functions of each criterion on interval scales, as in \cite{Corrente2021}. It has thus been used in this work with this dual purpose. The DCM, the associated notation, and other MCDA concepts used in this paper are defined in the next subsections, complemented with examples presented in Section \ref{subsection:valuefunctions} and \ref{appendix:a}.

\subsection{Basic data}
\label{subsection:basicdata}
\noindent The basic data of our problem is composed of the following elements:

\begin{itemize}[label={--}]
    \item $A = \{a_1,\ldots,a_i,\ldots,a_m\}$ is a finite the set of actions, in this case of ships, to be assessed and classified; this set is not necessarily known \textit{a priori}. 
    \item $G = \{g_1,\ldots,g_j,\ldots,g_n\}$ is the set of criteria do be considered for assessing the ships; some criteria can be considered as acceptation/rejection criteria. 
    \item $g_j(a_i)$ is the performance of action $a_i \in A$ on criteria $g_j$, $j=1,\ldots,n$.  
    \item $E_j = \{l_{j,1},\ldots,l_{j,k},\ldots,l_{j,t}\}$ is the scale of criterion $g_j$, in which $l_{j,k}$ represents a level on a discrete scale; in case of a continuous scale a continuous interval is used instead.
    \item $C = \{C_1,\ldots,C_r,\ldots,C_s\}$ is a set of totally ordered categories from the best, $C_1$, to the worst, $C_s$: $C_1 \succ \cdots \succ C_r \succ \cdots \succ C_s$, where $\succ$ means ``strictly preferred to''.  
\end{itemize}

\subsection{Aggregation model}
\label{subsection:aggregationmodel}
\noindent The problems dealt by MCDA involve the aggregation of the criteria, $G$, for the assessment of actions \citep{Roy1999}. This is achieved through \textit{aggregation models}, which make use of procedures to aggregate the partial information on the different criteria into an aggregated or comprehensive information. Three main types of aggregation models can be identified \citep{Figueira2016}: 1) value-/utility-based models; 2) outranking-based models; and 3) rule-based systems. In the current study, an additive value model is proposed as a complement to a rule-based system for the assignment of ships to risk categories, which will be described in the following paragraphs.

\paragraph{A. The additive model}
When facing a choosing, ranking, or sorting problem, an additive model allows a DM to optimize her/his decision, by showing which action maximizes the \textit{utility} (or \textit{value}) \citep{Keeney1993}. The DM preference system can be modelled through a comprehensive binary relation, $\succsim$, whose meaning is ``at least as good as'', over the set of actions to be compared, $A$. Thus, an action $a^{\prime}$ is considered to be as good as an action $a^{\prime\prime}$, denoted $a^{\prime} \succsim a^{\prime\prime}$, if and only if, the comprehensive value of $a^{\prime}$, $v(a^{\prime})$ is greater than of equal, the comprehensive value of $a^{\prime\prime}$, $v(a^{\prime\prime})$, i.e., $v(a^{\prime}) \geqslant v(a^{\prime\prime})$, where the comprehensive value of each action is additively computed as follows: 

\begin{equation}
\label{eq:1}
    v(a) = \sum_{j=1}^{n}w_jv_j\big(g_j(a)\big), \;\, \mbox{for all}\;\, a \in A
\end{equation}

\noindent in which $w_j$ are the criteria weights, for $j=1,\ldots,n$, (assuming that $\sum_{j=1}^{n}w_j =1$), and $v_j\big(g_j(a)\big)$ is the value of the performance $a$ on criterion $g_j$, $g_j(a)$, for all for $j=1,\ldots,n$.

\paragraph{B. Building the value functions (interval scales)}
The construction of the partial value functions, $v_j(\cdot)$, for $j=1,\ldots,n$ is performed in this study according to the DCM, as proposed in \cite{Corrente2021}. The scores $v_j(\cdot)$ are values in the interval scales \citep{Roberts1985}, built from the criteria scales, which can be discrete or continuous.

\paragraph{B.1. Discrete scales}
For a discrete criterion scale, of the type $E_j = \{l_{j,1},\ldots,l_{j,k},\ldots,l_{j,t}\}$, it is possible to determine the values of each scale level, i.e., $v_j(E_j) = \left\{v_j(l_{j,1}),\ldots,v_j(l_{j,k}),\ldots,v_j(l_{j,t})\right\}$, for each criterion $j=1,\ldots,n$, through the DCM. The procedure, which involves the elicitation of preferences from the DM, is shown through an example for one criterion in Section \ref{subsubsection:g1}, and later in \ref{appendix:a} for the remainder criteria modelled through discrete scales.

\paragraph{B.2. Continuous scales}
For a continuous criterion scale, a set of breakpoints of the scale, $E_j = [l_{j,1},l_{j,2}] \cup \cdots \cup [l_{j,k-1},l_{j,k}] \cup \cdots \cup [l_{j,t-1},l_{j,t}]$, are used to build a piecewise linear value function, to which the values $v_j(E_j) = [v_j(l_{j,1}),v_j(l_{j,2})] \cup \cdots \cup [v_j(l_{j,k-1}),v_j(l_{j,k})] \cup \cdots \cup [v_j(l_{j,t-1}),v_j(l_{j,t})]$ are assigned through a process similar to that used with discrete scales. Values for the real numbers in between the breakpoints can then be obtained through linear interpolation. An example is presented in Section \ref{subsubsection:g2}, referring to the only criterion modelled through a continuous scale in this study.

\paragraph{C. Determining the weights of criteria (ratio scales)}
In MCDA, the meaning of \textit{criteria weights} depends on the methodology being used. It may represent ``relative importance'' as in outranking methods \citep{Figueira2016}, ``priorities ratio'' as in the Analytic Hierarchy Process (AHP) \citep{Saaty1977}, or ``substitution rates'' or ``scale factors'' as in multiattribute utility theory (MAUT) methods \citep{Keeney1993}. The criteria weights assume the latter meaning in the present study. \textit{Substitution rates} refer to trade-offs in which the increase on a given criterion value is compensated, proportionally, by the decrease on the value of another criterion, or criteria \citep{Martel2016}. The process for assigning weights to the criteria used in this work follows that presented in \cite{Corrente2021}, in which \textit{dummy projects}, in this case \textit{dummy ships}, are defined and ranked. Similarly to the ``swing weights'' procedure established in \cite{VonWinterfeldt1986}, each dummy ship is defined by a ``swing'' in a given criterion, i.e., in each dummy ship, a criterion is changed from a reference level in the worst part of the scale (not necessarily the worst level of the scale) to a reference level in the best part of the scale (not necessarily the best level of the scale), keeping all the remainder criteria at their worst reference levels. Then, the DM ranks the dummy ships according to her/his preference, consequently ranking the different criteria. Finally, the \textit{closeness} between the criteria weights shall be assessed. Instead of the direct elicitation of the  $z$-ratio used in \cite{Corrente2021}, this work exploits the indifference felt by the DM between the dummy ship in the fist position in the ranking and the dummy ship in the worst position in the same ranking. The procedure followed for the criteria weight assignment is presented in Section \ref{subsection:weights}.

\subsection{Classification system}
\label{subsection:classificationsystem}
\noindent As previously mentioned, the purpose of the SRP is to assign risk categories to ships according to their performance in a defined set of criteria. In MCDA, this represents a sorting problem. The classification system followed in this study, i.e., the system through which the different risk categories are assigned to ships, is a hybrid model, which attends to the principles of the current Paris MoU SRP, namely by using the same risk categories (from the best to worst): Low Risk Ship (LRS), category $C_1$; Standard Risk Ship (SRS), category $C_2$; and High Risk Ship (HRS), category $C_3$. For the lowest risk category, $C_1$, a rule-based system is used, filtering the ships that can be considered as LRS. For the medium and high risk categories, $C_2$ and $C_3$, respectively, a value-based system is used instead. The defined risk categories are presented in Section \ref{subsection:definitionofcategories} and the results of the application of the developed hybrid system to the SRP are presented in Section \ref{subsection:resultsanddiscussion}.

\section{Application of the DCM for ship risk classification}
\label{section:application}
\noindent In this Section, the DCM is used in the construction of an MCDA model, proposed as an alternative to the Paris MoU SRP. First, a framework for the considered criteria is defined. The framework is then used in a sample of ships to establish their performances on the criteria. Next, the criteria value functions are built, the criteria weights are elicited, and the model categories are defined. Finally, the results obtained for the considered data sample are presented and discussed, including the presentation of a robustness analysis on the model's parameters.

\subsection{Construction of the criteria}
\label{subsection:constructionofcriteria}
\noindent The criteria used in this study are based on those established in the Paris MoU SRP \citep{ParisMoU2020}. Nonetheless, the criteria have been organized in a framework established with the purpose of facilitating the assessment and comparison of ships, as well as the communication of the decision model and its results. The terms used are based on those defined in \cite{Roy1999}. The ``Points of View'' constitute the upper level of the framework. A \textit{point of view} (PV) represents a more or less open angle under which each ship can be observed. Each point of view is characterized in terms of ``Significance Axes'', an intermediate level of the framework. A \textit{significance axis} (SA) is an area of analysis of each ship, to which a criterion refers to. Finally, each significance axis comprises at least one criterion, the lowest level of the framework. As mentioned previously in \Cref{section:methodology}, a \textit{criterion} is a model constructed for assessing the performances of the ships, allowing for their comparison. The established framework is presented next. 

\begin{enumerate}
    \item Point of View \textit{``Ship Characteristics and History''} (PV-SC\&H): This PV observes each ship by its \textit{intrinsic} aspects presented in the Paris MoU SRP, namely its physical properties and history. It is characterised through two Significance Axes: 
    \begin{enumerate}[label=(\roman*)]
        \item Significance Axis \textit{``Ship Characteristics''} (SA-CHAR): This SA analyses each ship through its physical properties. The SA is operationalised through the following criteria: 
        
        \begin{enumerate}[label=(\alph*)]
            \item \underline{Ship accident consequences}. This criterion is used to model and assess each ship according to the potential consequences in terms of societal, environmental, and financial losses, or human casualties, that, in case of accident or incident, might result from a ship of its type. 
            \begin{itemize}[label={--}]
                \item Code: ACCI. Notation: $g_{1}$. Preference direction: Minimization. 
                \item Scale type: Direct (not proxy), discrete, qualitative (purely ordinal). Scale unit: Verbal statements.
                \item Scale levels:
                \begin{itemize}
                    \item[$\circ$]  $l_{1,1}$ (low): ship types whose consequences, in case of accident or incident, are expected to be less significant;
                    \item[$\circ$]  $l_{1,2}$ (high): ship types including chemical tankers, gas carriers, oil tankers, bulk carriers, passenger ships, and NLS tankers, whose consequences, in case of accident or incident, are expected to be more significant.
                \end{itemize}
                \item Comments:  The scale associated with this criterion contains the performance levels of different ship types, as defined in the SRP.
            \end{itemize}
            \item \underline{Age of ship}. This criterion is used to model and assess each ship according to its age.
            \begin{itemize}[label={--}]
                \item Code: AGES. Notation: $g_{2}$. Preference direction: Minimization.
                \item Scale type: Direct, continuous, quantitative. Scale unit: Numerical values.
                \item Scale levels: Real values
                \item Comments: Instead of a discrete scale, such as that used in the Paris MoU, a continuous quantitative scale is used in this work. The scale associated with this criterion models the performance of ships with different ages.
            \end{itemize}
        \end{enumerate}
        \item Significance Axis \textit{``Ship History''} (SA-HIST): This SA analyses each ship through its historical records. The SA is operationalised through the following criteria: 
        \begin{enumerate}[label=(\alph*)]
            \item \underline{Deficiencies}. This criterion is used to model and assess each ship according to the number of recorded deficiencies.
            \begin{itemize}[label={--}]
                \item Code: DEFC. Notation: $g_{3}$. Preference direction: Minimization.
                 \item Scale type: Direct, discrete, qualitative (purely ordinal). Scale unit: Verbal statements.
                \item Scale levels:  
                \begin{itemize}
                    \item[$\circ$]  $l_{3,1}$ (low): ships with 5 deficiencies or fewer in the last 36 months, with at least one inspection performed;
                    \item[$\circ$]  $l_{3,2}$ (medium): ships with more than 5 deficiencies recorded in the last 36 months, with at least one inspection performed;
                    \item[$\circ$]  $l_{3,3}$ (high): ships that are not eligible, i.e. ships that have not performed at least one inspection in the last 36 months. 
                \end{itemize}
                \item Comments: The scale associated with this criterion contains the performance levels of ships with different number of recorded deficiencies, as established in the SRP.
            \end{itemize}
            \item \underline{Detentions}. This criterion is used to model and assess each ship according to the number of recorded detentions.
            \begin{itemize}[label={--}]
                \item Code: DETN. Notation: $g_{4}$. Preference direction: Minimization.
                \item Scale type: Direct, discrete, qualitative (purely ordinal). Scale unit: Verbal statements.
                \item Scale levels: 
                \begin{itemize}
                    \item[$\circ$]  $l_{4,1}$ (no): ships with no detentions in the last 36 months;
                    \item[$\circ$]  $l_{4,2}$ (one): ships with one detention in the last 36 months;
                    \item[$\circ$]  $l_{4,3}$ (more): ships with two detentions or more in the last 36 months.
                \end{itemize}
                \item Comments: The scale associated with this criterion contains the performance levels of ships with different number of recorded detentions, as established in the SRP.
            \end{itemize}
        \end{enumerate}
    \end{enumerate}
    \item Point of View \textit{``Ship Registration and Classification''} (PV-SR\&C): This PV observes each ship by its \textit{extrinsic} aspects presented in the Paris MoU SRP, namely the legal requirements that it has to comply with in order to operate. The PV is defined over three Significance Axes:
    \begin{enumerate}[label=(\roman*)]
        \item Significance Axis \textit{``Ship Company''} (SA-COMP): This SA analyses each ship through the company responsible for its compliance with the International Safety Management (ISM) Code.\footnote{See \url{https://www.imo.org/en/OurWork/HumanElement/Pages/ISMCode.aspx}.} The SA is operationalised through the following criterion: 
        \begin{enumerate}[label=(\alph*)]
            \item \underline{Company performance}. This criterion is used to model and assess each ship according to the ISM Company performance.
            \begin{itemize}[label={--}]
                \item Code: COPF. Notation: $g_{5}$. Preference direction: Maximization.
                \item Scale type: Direct, discrete, qualitative (purely ordinal). Scale unit: Verbal statements.
                \item Scale levels: 
                \begin{itemize}
                    \item[$\circ$]  $l_{5,1}$ (low): companies whose performance is defined as ``very low'' or ``low'';
                    \item[$\circ$]  $l_{5,2}$ (medium): companies whose performance is defined as ``medium'';
                    \item[$\circ$]  $l_{5,3}$ (high): companies whose performance is defined as ``high''.
                \end{itemize}
                \item Comments: The scale associated with this criterion contains the performance levels of ships with different company performances, as established in the SRP. 
            \end{itemize}
        \end{enumerate}
        \item Significance Axis \textit{``Ship Flag State''} (SA-FLAG): This SA analyses each ship through the flag State under which it is registered. The SA is operationalised through the following criteria:
        \begin{enumerate}[label=(\alph*)]
            \item \underline{Flag performance}. This criterion is used to model and assess each ship according to the flag State performance. 
            \begin{itemize}[label={--}]
                \item Code: FLPF. Notation: $g_{6}$. Preference direction: Maximization.
                \item Scale type: Direct, discrete, qualitative (purely ordinal). Scale unit: Verbal statements.
                \item Scale levels: 
                \begin{itemize}
                    \item[$\circ$]  $l_{6,1}$ (very low): flag States identified as ``Black'' with ``very high risk'', ``high risk'', or ``medium to high risk'';
                    \item[$\circ$]  $l_{6,2}$ (low): flag States identified as ``Black'' with ``medium risk'';
                    \item[$\circ$]  $l_{6,3}$ (medium): flag States identified as ``Grey'' or not listed in the ``BGW list'';
                    \item[$\circ$]  $l_{6,4}$ (high): flag States identified as ``White''. 
                \end{itemize}
                \item Comments: The flag State performance is established annually by the Paris MoU Committee in the ``BGW list'' (Black, Grey, and White). The scale associated with this criterion contains the performance levels of ships with different flag State performances, as established in the SRP.
            \end{itemize}
            \item \underline{Fulfilment of the International Maritime Organisation Audit}. This criterion is used to model and assess each ship according to the fulfilment by its flag State of the requirement regarding the presentation of the IMO Audit (IA) report.
            \begin{itemize}[label={--}]
                \item Code: FLIA. Notation: $g_{7}$. Preference direction: Maximization.
                \item Scale type: Direct, discrete, qualitative (purely ordinal). Scale unit: Verbal statements.
                \item Scale levels: 
                \begin{itemize}
                    \item[$\circ$]  $l_{7,1}$ (no): flag States which do not fulfil the IA report requirement;
                    \item[$\circ$]  $l_{7,2}$ (yes): flag States which fulfil such requirement.
                \end{itemize}
                \item Comments: The list of flag States fulfilling the requirement regarding the presentation of the IA report is published annually by the Paris MoU Committee. The scale associated with this criterion differentiates ships between those registered to flag States that fulfil the requirement and those registered to flag States that do not.
            \end{itemize}
        \end{enumerate}
        \item Significance Axis \textit{``Recognised Organisation''} (SA-RECO): This SA analyses each ship through its Recognised Organisation (RO).\footnote{See \url{https://www.imo.org/en/OurWork/MSAS/Pages/RecognizedOrganizations.aspx}.} The SA is operationalised through the following criteria:
        \begin{enumerate}[label=(\alph*)]
            \item \underline{Recognised Organisation performance}. This criterion is used to model and assess each ship according to the RO performance.
            \begin{itemize}[label={--}]
                \item Code: ROPF. Notation: $g_{8}$. Preference direction: Maximization.
                \item Scale type: Direct, discrete, qualitative (purely ordinal). Scale unit: Verbal statements.
                \item Scale levels: 
                \begin{itemize}
                    \item[$\circ$]  $l_{8,1}$ (low): ROs whose performance is defined as ``very low'' or ``low'';
                    \item[$\circ$]  $l_{8,2}$ (medium): ROs whose performance is defined as ``medium'';
                    \item[$\circ$]  $l_{8,3}$ (high): ROs whose performance is defined as ``high''.
                \end{itemize}
                \item Comments: The RO performance is established annually by the Paris MoU Committee in the ``RO Performance list''. The scale associated with this criterion contains the performance levels of ships with different RO performances, as established in the SRP.
            \end{itemize}
            \item \underline{Recognition by at least one member State of the Recognised Organisation}. This criterion is used to model and assess each ship according to the recognition of its RO by at least one member State.
            \begin{itemize}[label={--}]
                \item Code: RORE. Notation: $g_{9}$. Preference direction: Maximization.
                \item Scale type: Direct, discrete, qualitative (purely ordinal). Scale unit: Verbal statements.
                \item Scale levels: 
                \begin{itemize}
                    \item[$\circ$]  $l_{9,1}$ (no): ROs not recognised by at least one member State;
                    \item[$\circ$]  $l_{9,2}$ (yes): ROs recognised by one member State or more.
                \end{itemize}
                \item Comments: The list of ROs recognised by at least one member State is published by the Paris MoU Committee. The scale associated with this criterion differentiates ships between those with ROs that fulfil this requirement and those without.
            \end{itemize}
        \end{enumerate}
    \end{enumerate}
\end{enumerate}

\subsection{Performance Table}
\label{subsection:definitionofperformances}
\noindent This study uses a data sample of 136 ships, inspected a total of 138 times at the port of Lisbon, Portugal, through PSC inspections. The period under analysis ranges from January 1 to December 31, 2018, and the data sample has been collected from the THETIS platform.\footnote{Available at \url{https://portal.emsa.europa.eu/web/thetis/inspections}.} A set of ten ships from the collected sample is presented in \Cref{table:ships}. The ISM company names have been coded for privacy reasons.
\begin{table}[H]
\small
\centering
\caption{Data sample.}
\label{table:ships}
\begin{center}
\begin{tabular}{|c |c c c c c c c|} 
    \hline 
    \multirow{2}{*}{Ship} & \multirow{2}{*}{Type} & \multirow{2}{*}{Age} & \multirow{2}{*}{Deficiencies} & \multirow{2}{*}{Detentions} & \multirow{2}{*}{ISM Company} & \multirow{2}{*}{Flag} & \multirow{1}{*}{Recognised} \\
    & & & & & & & \multirow{1}{*}{Organisation} \\
    \hline
    $a_{1}$ & Refrig. cargo & 18 & 2 & 0 & ISM 12 & Italy & RINA \\
    $a_{2}$ & Container & 17 & 3 & 0 & ISM 55 & Honk Kong & DNVGL \\ 
    $a_{3}$ & Container & 7& 11 & 0 & ISM 110 & Germany & DNVGL \\ 
    $a_{4}$ & Bulk carrier & 2 & 0 & 0 & ISM 107 & Panama & NKK \\
    $a_{5}$ & Container & 10 & 4 & 0 & ISM 5 & Cyprus & DNVGL \\ 
    $a_{6}$ & Bulk carrier & 22 & 15 & 0 & ISM 45 & Liberia & NKK \\
    $a_{7}$ & Bulk carrier & 11 & 4 & 0 & ISM 71 & Italy & RINA \\
    $a_{8}$ & Bulk carrier & 15 & 10 & 0 & ISM 19 & Panama & NKK \\
    $a_{9}$ & General cargo & 28 & 0 & 0 & ISM 24 & Barbados & BV \\
   $a_{10}$ & Oil tanker & 11 & 0 & 0 & ISM 3 & Singapore & ABS \\
    \hline
\end{tabular}
\end{center}
\end{table}

The performance table resulting from the application of the criteria developed previously to the set of ships from \Cref{table:ships} is presented in \Cref{table:performances}.
\begin{table*}[htp!]
\small
\def\arraystretch{1.5}
\newcolumntype{L}[1]{>{\raggedright\let\newline\\\arraybackslash\hspace{0pt}}m{#1}}
\newcolumntype{C}[1]{>{\centering\let\newline\\\arraybackslash\hspace{0pt}}m{#1}}
\centering
\caption{Performance Table.}
\label{table:performances}
\resizebox{\textwidth}{!}{
\begin{tabularx}{1\textwidth} { 
    | >{\centering\arraybackslash}X
    | >{\centering\arraybackslash}X
     >{\centering\arraybackslash}X
     >{\centering\arraybackslash}X
     >{\centering\arraybackslash}X
     >{\centering\arraybackslash}X
     >{\centering\arraybackslash}X
     >{\centering\arraybackslash}X
     >{\centering\arraybackslash}X
     >{\centering\arraybackslash}X | }
    \hline
    \multirow{5}{*}{Ship} & \multicolumn{4}{c|}{\sc{pv-sc\&h}} & \multicolumn{5}{c|}{\sc{pv-sr\&c}} \\
    \cline{2-10}
    & \multicolumn{2}{c|}{\sc{sa-char}}  & \multicolumn{2}{c|}{\sc{sa-hist}} & \multicolumn{1}{c|}{\sc{sa-comp}} & \multicolumn{2}{c|}{\sc{sa-flag}} & \multicolumn{2}{c|}{\sc{sa-reco}} \\
    \cline{2-10}
    & \multicolumn{1}{c|}{\sc{acci}} & \multicolumn{1}{c|}{\sc{ages}} & \multicolumn{1}{c|}{\sc{defc}} & \multicolumn{1}{c|}{\sc{detn}} & \multicolumn{1}{c|}{\sc{copf}} & \multicolumn{1}{c|}{\sc{flpf}} & \multicolumn{1}{c|}{\sc{flia}} & \multicolumn{1}{c|}{\sc{ropf}} & \multicolumn{1}{c|}{\sc{rore}} \\
    \cline{2-10}
    & \multicolumn{1}{c|}{$g_{1}$} & \multicolumn{1}{c|}{$g_{2}$} & \multicolumn{1}{c|}{$g_{3}$} & \multicolumn{1}{c|}{$g_{4}$} & \multicolumn{1}{c|}{$g_{5}$} & \multicolumn{1}{c|}{$g_{6}$} & \multicolumn{1}{c|}{$g_{7}$} & \multicolumn{1}{c|}{$g_{8}$} & \multicolumn{1}{c|}{$g_{9}$} \\
    \cline{2-10}
    & \multicolumn{1}{c|}{$min$} & \multicolumn{1}{c|}{$min$} & \multicolumn{1}{c|}{$min$} & \multicolumn{1}{c|}{$min$} & \multicolumn{1}{c|}{$max$} & \multicolumn{1}{c|}{$max$} & \multicolumn{1}{c|}{$max$} & \multicolumn{1}{c|}{$max$} & \multicolumn{1}{c|}{$max$} \\
    \hline
    $a_{1}$ & low & 18 & low & no & medium & high & yes & high & yes \\
    $a_{2}$ & low & 17 & low & no & medium & high & yes & high & yes \\
    $a_{3}$ & low & 7 & medium & no & medium & high & yes & high & yes \\
    $a_{4}$ & high & 2 & low & no & high & high & yes & high & yes \\
    $a_{5}$ & low & 10 & low & no & medium & high & yes & high & yes \\
    $a_{6}$ & high & 22 & medium & no & low & high & yes & high & yes \\
    $a_{7}$ & high& 11 & low & no & medium & high & yes & high & yes \\
    $a_{8}$ & high & 15 & medium & no & medium & high & yes & high & yes \\
    $a_{9}$ & low & 28 & low & no & medium & high & no & high & yes \\
    $a_{10}$ & high & 11 & low & no & high & high & yes & high & yes \\
    \hline
\end{tabularx}
}
\end{table*}
\subsection{Criteria value functions}
\label{subsection:valuefunctions}
\noindent After the definition of the criteria framework presented in Section \ref{subsection:constructionofcriteria}, the DCM, as established in \cite{Corrente2021}, has been used for the construction of value functions for the criteria. This method requires an interaction between an analyst, or analysts team, and a single DM, or multiple DMs. In the case of this work, the analysts team consists of the authors, while the DM is an expert with a research background on maritime policy, particularly on the study of the Paris MoU. 

Pairwise comparisons have been performed between the scale levels of each criterion. The numbers in the comparison tables, such as the one presented in \Cref{table:g2}, represent blank cards inserted between any two levels, which, in turn, models the preference difference between the levels. In this Section, criterion $g_1$ (ship accident consequences) and criterion $g_2$ (age of ship) are presented as examples on the construction of the value functions, as the former comprises only two scale levels, and the latter comprises multiple levels. The same process has been used for the value functions of criteria $g_{3}$ (deficiencies), $g_{4}$ (detentions), $g_{5}$ (company performance), $g_{6}$ (flag performance), and $g_{8}$ (RO performance), presented in \Cref{fig:criteriascales}. Criteria $g_{7}$ (IMO audit) and $g_{9}$ (RO recognised) do not have an associated value function as these are acceptation/rejection criteria. Instead, they are used in Section \ref{subsection:definitionofcategories} as part of the rule based filtering system.

\subsubsection{Criterion $g_1$: \textit{Ship accident consequences}}
\label{subsubsection:g1}
\begin{figure}[htp!]
    \centering
    \begin{subfigure}[b]{0.3\textwidth}
        \centering
        \includegraphics[width=\textwidth,height=4in]{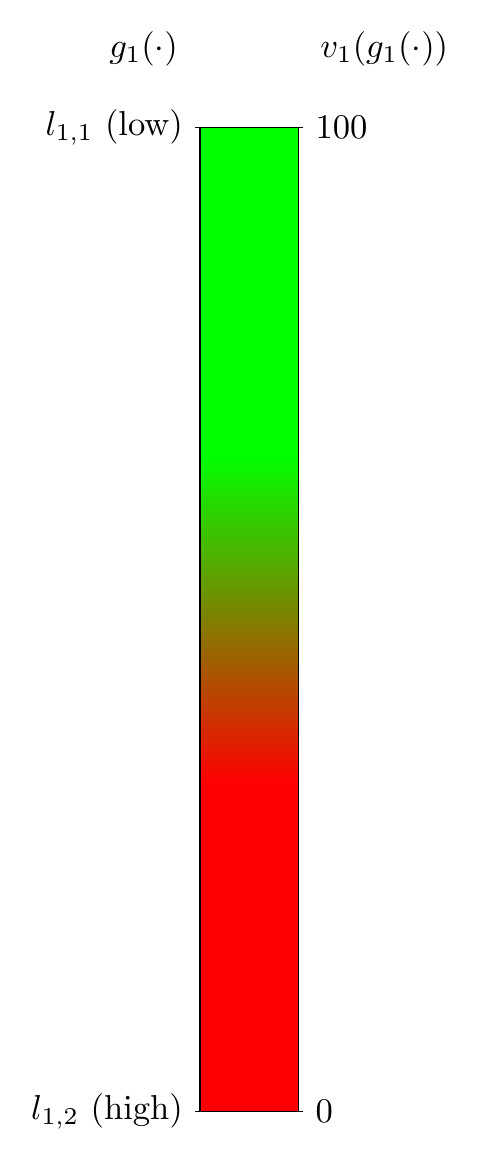}
        \caption{Criterion $g_1$ (ACCI).}    
        \label{criteriascales:a}
    \end{subfigure}
    \begin{subfigure}[b]{0.3\textwidth}  
        \centering 
        \includegraphics[width=\textwidth,height=4in]{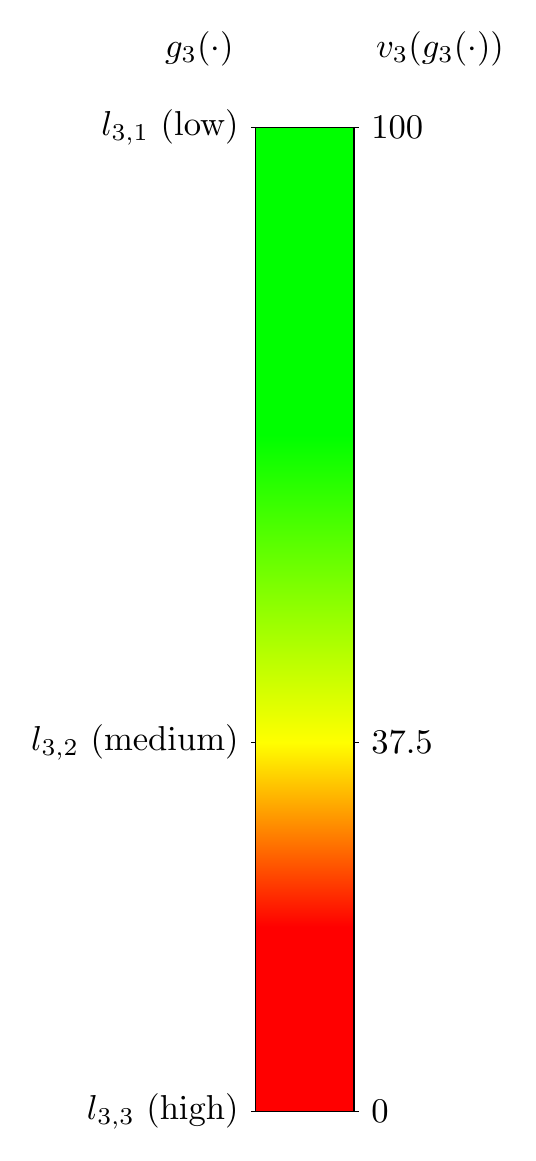}
        \caption{Criterion $g_3$ (DEFC).}    
        \label{criteriascales:b}
    \end{subfigure}
    \begin{subfigure}[b]{0.3\textwidth}  
        \centering 
        \includegraphics[width=\textwidth,height=4in]{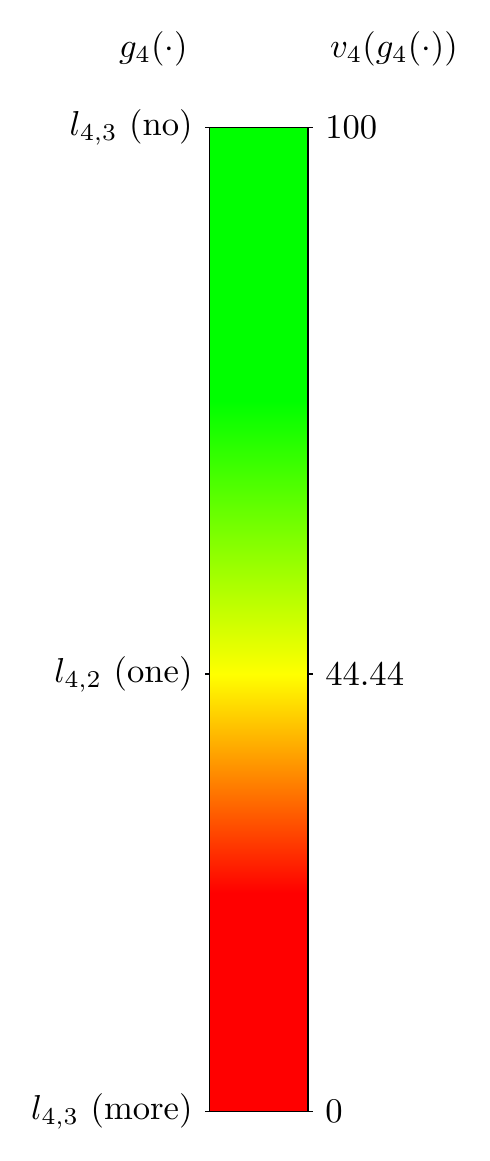}
        \caption{Criterion $g_4$ (DETN).}    
        \label{criteriascales:c}
    \end{subfigure}
    
    \bigskip
    
    \centering
    \begin{subfigure}[b]{0.3\textwidth}
        \centering
        \includegraphics[width=\textwidth,height=4in]{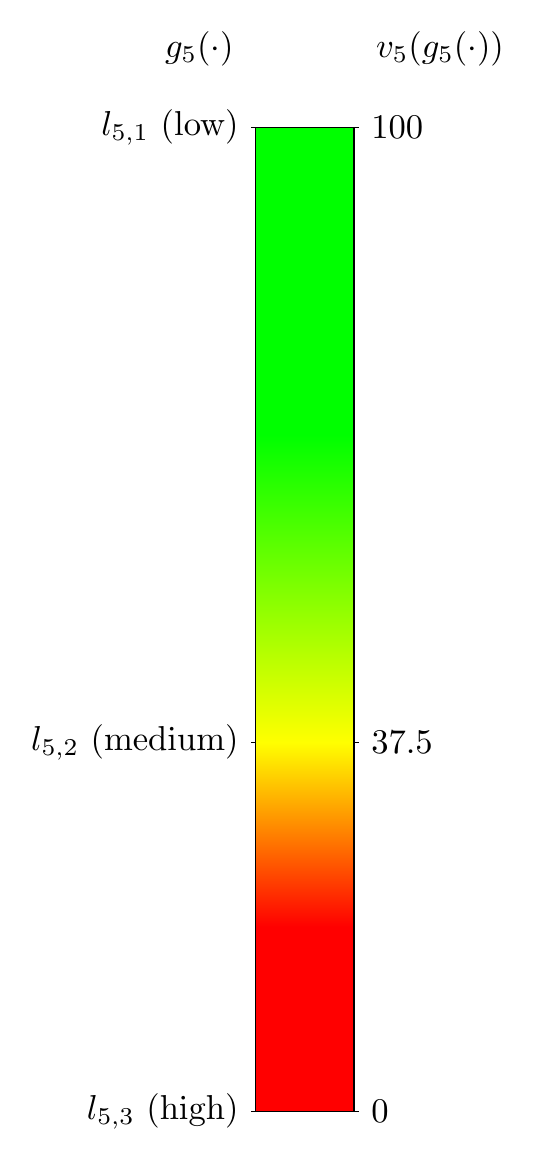}
        \caption{Criterion $g_5$ (COPF).}    
        \label{criteriascales:d}
    \end{subfigure}
    \begin{subfigure}[b]{0.3\textwidth}  
        \centering 
        \includegraphics[width=\textwidth,height=4in]{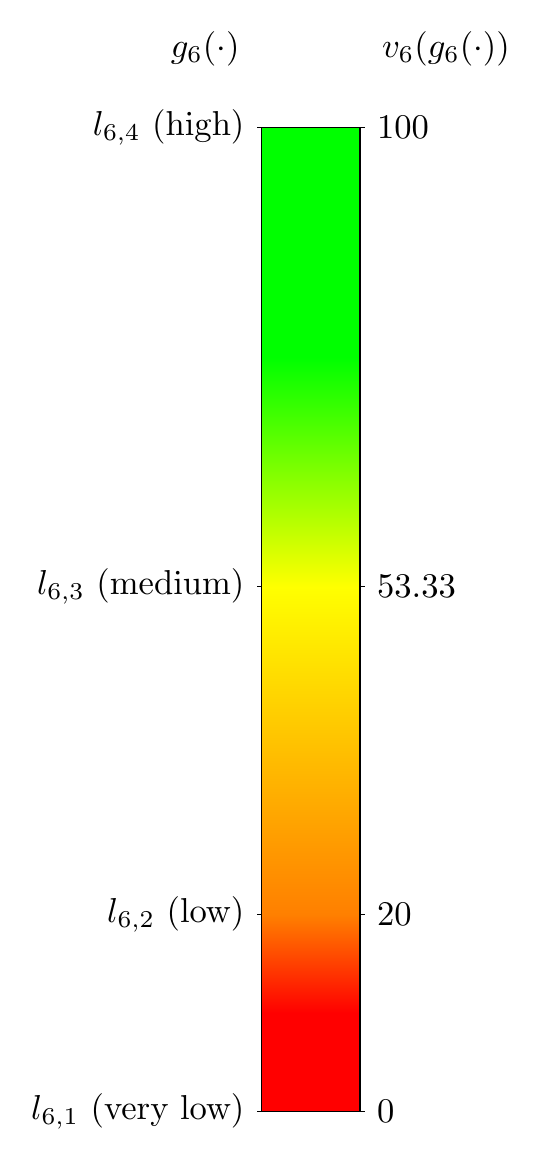}
        \caption{Criterion $g_6$ (FLPF).}    
        \label{criteriascales:e}
    \end{subfigure}
    \begin{subfigure}[b]{0.3\textwidth}  
        \centering 
        \includegraphics[width=\textwidth,height=4in]{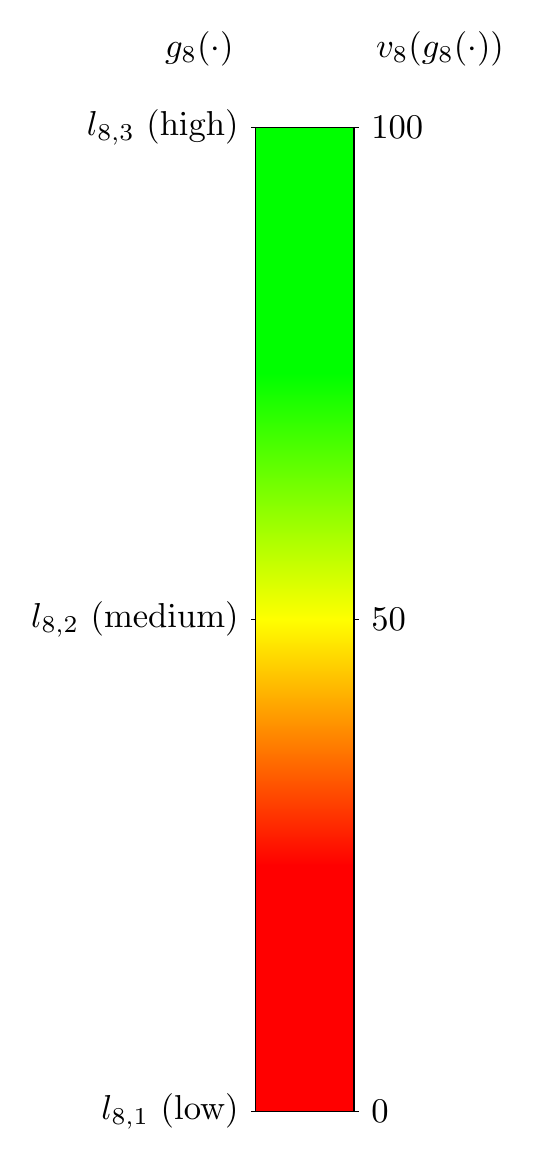}
        \caption{Criterion $g_8$ (ROPF).}    
        \label{criteriascales:f}
    \end{subfigure}
    
\caption{Criteria scales.} 
\label{fig:criteriascales}
\end{figure}

\noindent Criterion $g_1$ is modeled through a two-level scale: $l_{1,1}$, which refers to ship types non-listed in the SRP and whose consequences in case of accident are expected to be less significant; and $l_{1,2}$, which refers to  ship types listed in the SRP and whose consequences in case of accident are expected to be more significant. In this case, since only two levels have been considered for the scale, a value of 100 has been assigned to $l_{1,1}$ and a value of 0 has been assigned to $l_{1,2}$. The representation of the scale for criterion $g_1$ is presented in \Cref{criteriascales:a}.\\

\subsubsection{Criterion $g_2$: \textit{Age of ship}}
\label{subsubsection:g2}
\noindent Criterion $g_2$ refers to the age of the the ships, which is, naturally, defined over a continuous domain. In the Paris MoU SRP, however, ships are only differentiated between those that older than 12 years and those that are newer. In this study, the criterion scale has been modelled through a piecewise linear value function, whose breakpoints are based on ship ages used by the \cite{USCG2016} for their characterisation:
\begin{itemize}
    \item[$\circ$]  $l_{2,1}$: new ships (0 years old);
    \item[$\circ$]  $l_{2,2}$: ships 5 years old;
    \item[$\circ$]  $l_{2,3}$: ships 10 years old;
    \item[$\circ$]  $l_{2,4}$: ships 15 years old;
    \item[$\circ$]  $l_{2,5}$: ships 20 years old;
    \item[$\circ$]  $l_{2,6}$: ships 25 years old or older.
\end{itemize}
As previously stated, the values for these levels have been obtained through the DCM, as established in \cite{Corrente2021}. The following steps describe the interaction between the analysts team and the DM on the construction of the value function for criterion $g_2$. Value functions for the remainder criteria have been built similarly and the results are presented in \Cref{fig:criteriascales}.
\begin{enumerate}
    \item Each level was written on a card. Additional information may be included, if needed.
    \item In the case of criterion $g_2$, the scale levels, which refer to the breakpoints presented before, have the following order:
    \[
        l_{2,6} \prec l_{2,5} \prec l_{2,4} \prec l_{2,3} \prec l_{2,2} \prec l_{2,1}
    \]
    \item Two reference levels were identified by the DM. In this case, the reference levels are $l_{2,1}$ and $l_{2,6}$, which are assigned with the values $v_2(l_{2,1})=100$ and $v_2(l_{2,6})=0$, respectively. In other words, new ships have been assigned the maximum preference value (100), while ships with twenty-five years or more have been assigned the minimum preference value (0). Nonetheless, other levels could have been chosen by the DM as reference levels, and not necessarily the best and the worst.
    \item The DM was then asked to add blank cards (of which a sufficiently large number has been provided) between pairs of levels. According to the DCM, the blank cards are used to model the differences in preference (i.e., intensities of preferences) between pairs of levels, which in this case are the levels of the criteria scales. The number of blank cards refers to the following:
    \begin{itemize}[label={--}]
        \item Zero blank cards between a pair of levels does not mean that the two levels have the same value, but, instead, that the difference is minimal, equal to a unit, $\alpha$, whose value is computed as in Step 6 of the present procedure;
        \item One blank card means that the difference is twice the unit;
        \item Two blank cards mean that the difference is three times the unit;
        \item And so forth.
    \end{itemize}
    The differences in preference between scale levels for the age criterion have been collected from the DM and are presented in bold in \Cref{table:g2}. In this table only the preference differences for consecutive levels have been assessed (the diagonal of the table).
    \begin{table}[H]
    \small
    \centering
    \caption{Criterion $g_2$ (age of ship) comparison table.}
    \label{table:g2}
    \begin{center}
    \begin{tabular}{c |c c c c c c|} 
        \cline{2-7}
        & $l_{2,6}$ & $l_{2,5}$ & $l_{2,4}$ & $l_{2,3}$ & $l_{2,2}$ & $l_{2,1}$ \\
        \hline
        \multicolumn{1}{|l|}{$l_{2,6}$} & \cellcolor{black} & \textbf{0} & 3 & 7 & 11 & 16 \\ 
        \multicolumn{1}{|l|}{$l_{2,5}$} & & \cellcolor{black} & \textbf{2} & 6 & 10 & 15 \\ 
        \multicolumn{1}{|l|}{$l_{2,4}$} & & & \cellcolor{black} & \textbf{3} & 7 & 12 \\ 
        \multicolumn{1}{|l|}{$l_{2,3}$} & & & & \cellcolor{black} & \textbf{3} & 8 \\
        \multicolumn{1}{|l|}{$l_{2,2}$} & & & & & \cellcolor{black} & \textbf{4} \\ 
        \multicolumn{1}{|l|}{$l_{2,1}$} & & & & & & \cellcolor{black} \\ 
        \hline
    \end{tabular}
    \end{center}
    \end{table}
    From \Cref{table:g2} it can be observed that zero blank cards have been added by the DM between $l_{2,6}$ and $l_{2,5}$. In practice, the DM assessed the difference in preference between ships with twenty-five or more years and ships with twenty years as minimal, i.e., equal to $\alpha$. The difference between $l_{2,5}$ and $l_{2,4}$ has been assessed with two blank cards; The differences between $l_{2,4}$ and $l_{2,3}$, and $l_{2,3}$ and $l_{2,2}$ have been assessed with three blank cards each; Finally, the preference difference between $l_{2,2}$ and $l_{2,1}$ has been assessed with four blank cards, the largest difference between two consecutive levels in the comparison table of criterion $g_2$. This means that new ships are, not only the most preferred from the considered levels, but also more preferred than any other level in the pairwise comparisons.
    \item As established in \cite{Corrente2021}, more preference judgments can be obtained from the DM to fill the remaining elements of the table. However, in the present study this has been done by \textit{transitivity}. It means that the preference differences between non-consecutive levels are obtained following the consistency condition presented in \cite{Corrente2021}. The preference difference between two non-consecutive levels $p$ and $q$, $e_{pq}$, is obtained through:
    \begin{equation}
    \label{eq:2}
        e_{pq} = e_{pk} + e_{kq} + 1 \;\;\, \mbox{for all} \;\; p,k,q = 1,\ldots,t \quad \textrm{and} \quad p < k < q 
    \end{equation}
    For example, the difference between $l_{2,6}$ and $l_{2,4}$, $e_{64}$, is equal to: $e_{65} + e_{54} + 1 = 0 + 2 + 1 = 3$. The remaining preference values have been computed similarly.
    \item The value of the unit $\alpha$ has been computed from the values of the two reference levels, divided by the number of units between them:
    \begin{align*}
    \alpha = \frac{v_2(l_{2,6}) - v_2(l_{2,1})}{(0+1)+(2+1)+(3+1)+(3+1)+(4+1)} = \frac{100 - 0}{17} \approx 5.88
    \end{align*}
    \item The values of the remainder levels have been computed through $\alpha$:
    \[
        \left\{
        \begin{array}{l}
            v_2(l_{2,5}) = v_2(l_{2,6}) + (0 + 1) \times \alpha = 0 + (0 + 1) \times 5.88 = 5.88 \\
            v_2(l_{2,4}) = v_2(l_{2,6}) + (3 + 1) \times \alpha = 0 + (3 + 1) \times 5.88 = 23.52 \\
            v_2(l_{2,3}) = v_2(l_{2,6}) + (7 + 1) \times \alpha = 0 + (7 + 1) \times 5.88 = 47.04 \\
            v_2(l_{2,2}) = v_2(l_{2,6}) + (11 + 1) \times \alpha = 0 + (11 + 1) \times 5.88 = 70.56 \\
        \end{array}
        \right.
    \]
    Being a continuous scale, in opposition to the discrete scale used in the Paris MoU SRP, values for ships with ages between the considered levels can also be obtained. This is possible through linear interpolation between any two levels. For example, the value of a ship three years old on criterion $g_2$, $v_2(3)$, is equal to:
    \begin{align*}
    v_2(3) = v_2(l_{2,1}) - \frac{(v_2(l_{2,1}) - v_2(l_{2,2}))}{(l_{2,1} - l_{2,2})}(l_{2,1} - 3) = 100 - \frac{(100 - 70.56)}{(0 - 5)}(0 - 3) \approx 82.34
    \end{align*}
\end{enumerate}

\begin{figure}[H]
    \centering
    \includegraphics[width=10cm, height=7.5cm]{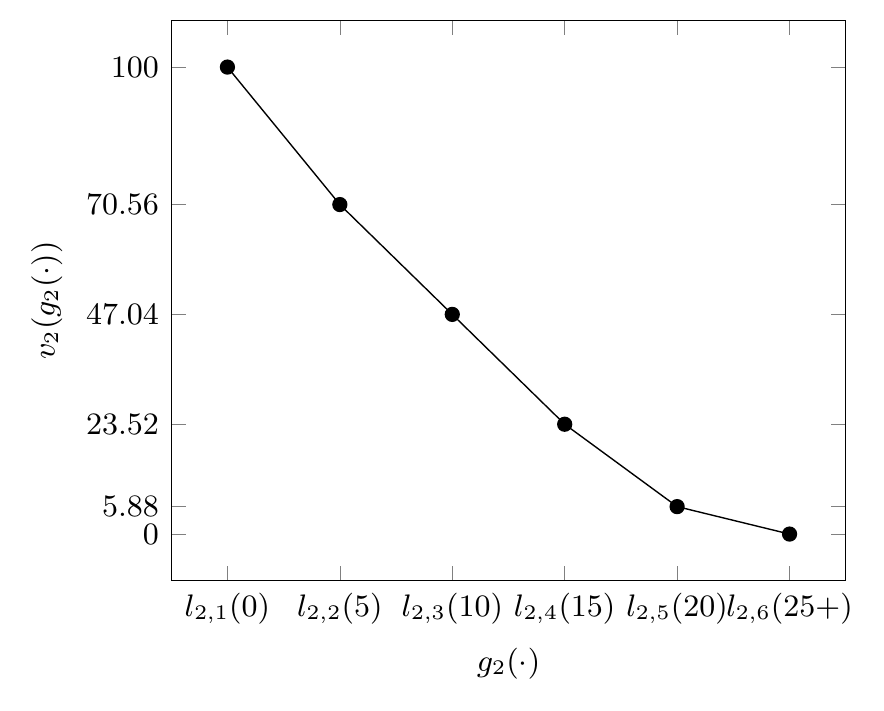}
    \caption{Criterion $g_2$ (age of ship) value function.}
    \label{fig:g2}
\end{figure}

The resulting value function for criterion $g_2$ is presented in \Cref{fig:g2}

\subsection{Criteria weights}
\label{subsection:weights}
\noindent  The questioning procedure for assessing the weights of criteria is similar to the one used for building the value functions, which, in turn, is similar to the procedure established in \cite{Corrente2021}. The steps presented below have been followed for determining the criteria weights:

\begin{enumerate}
    \item For each criterion, two reference levels have been considered, one in the worst part of the scale, ${g}^{r}_j$, and the other in the best part of the scale, ${g}^{s}_j$, together with the respective values, $v_j({g}^{r}_j)$ and $v_j({g}^{s}_j)$, for $j=1,\ldots,n$. In the case of this study, the performance levels and their respective values are presented in \Cref{table:performancelevels}.
    \begin{table}[H]
    \small
    \centering
    \caption{Performance levels and values.}
    \label{table:performancelevels}
    \begin{center}
    \begin{tabular}{|c|c c c c c c c|} 
        \hline
        Criterion & $g_{1}$ & $g_{2}$ & $g_{3}$ & $g_{4}$ & $g_{5}$ & $g_{6}$ & $g_{8}$ \\
        \hline
        Pref. direction & $\min$ & $\min$  & $\min$  & $\min$  & $\max$  & $\max$ & $\max$ \\
        ${g}^{r}_j$ & high & 25+ & high & more & low & very low & low \\
        ${g}^{s}_j$ & low & 0 & low & no & high & high & high \\ 
        $v_j({g}^{r}_j)$ & 0 & 0 & 0 & 0 & 0 & 0 & 0 \\ 
        $v_j({g}^{s}_j)$ & 100 & 100 & 100 & 100 & 100 & 100 & 100 \\ 
        \hline
    \end{tabular}
    \end{center}
    \end{table}
    \item Dummy actions have been built such that action $a_i$, in this case ship $s_i$, has the highest evaluation on criterion $g_j$ and the lowest on the remaining ones. In this study, seven dummy ships have been built as follows:
    \begin{itemize}[label={--}]
        \item $s_1 = (\mathbf{low},25+,high,more,low,very \; low,low) \equiv (\mathbf{100},0,0,0,0,0,0)$ 
        \item $s_2 = (high,\mathbf{0},high,more,low,very \; low,low) \equiv (0,\mathbf{100},0,0,0,0,0)$
        \item $s_3 = (high,25+,\mathbf{low},more,low,very \; low,low) \equiv (0,0,\mathbf{100},0,0,0,0)$
        \item $s_4 = (high,25+,high,\mathbf{no},low,very \; low,low) \equiv (0,0,0,\mathbf{100},0,0,0)$
        \item $s_5 = (high,25+,high,more,\mathbf{high},very \; low,low) \equiv (0,0,0,0,\mathbf{100},0,0)$
        \item $s_6 = (high,25+,high,more,low,\mathbf{high},low) \equiv (0,0,0,0,0,\mathbf{100},0)$
        \item $s_8 = (high,25+,high,more,low,very \; low,\mathbf{high}) \equiv (0,0,0,0,0,0,\mathbf{100})$
    \end{itemize}
    On each dummy ship a criterion is changed from the worst reference level to the best reference level, i.e., a ``swing'' \citep{VonWinterfeldt1986} is performed, while the remaining reference levels were kept at the worst reference level. In this case, the worst and best performances, as well as their values ($0$ and $100$, respectively), have been used as reference levels, although other references can be considered, depending on the problem.
    \item The dummy ships have been ranked by considering the \textit{swing} from ${g}^{r}_j$ to ${g}^{s}_j$, or, equivalently, from $v({g}^{r}_j)$ to $v({g}^{s}_j)$, for $j=1,\ldots,n$. Some actions can occupy the same position in the ranking, meaning that they are assigned the same weight. In the present case, the DM provided the following ranking for the considered dummy ships. 
    \[
        s_3 \prec s_4 \prec s_6 \prec s_1 \prec s_8 \prec s_5 \prec s_2
    \]
    The ranking provided by the DM means that a ship with the best reference performance on the criterion $g_2$ (age of ship), and the worst reference performance on the remainder criteria, is strictly preferred to a ship with the best reference performance on the criterion $g_5$ (company performance), and the worst reference performance on the remainder criteria, and so forth.
    
    After ranking the dummy ships, the DCM has been used to model their greater or lesser \textit{closeness} in terms of weights. In the present context, the weights represent \textit{substitution rates}, i.e., trade-offs in which the increase on a given criterion (or dummy ship) value is compensated, proportionally, by the decrease on the value of another criterion \citep{Martel2016}. There are different ways of assessing such a piece of information from the DM. In this case, the first dummy ship in the ranking provided in the last step, $s_2$, is considered the reference criterion, i.e., the criterion with the highest weight. Then, the DM has been asked to add a large enough number of blank cards, not between pairs of elements, as in Section \ref{subsection:valuefunctions}, but between the reference criterion (dummy ship $s_2$) and each of the remainder to measure their \textit{closeness}. The obtained judgments are presented in bold in \Cref{table:weights}.
    \begin{table}[H]
    \small
    \centering
    \caption{Dummy ships comparison table.}
    \label{table:weights}
    \begin{center}
    \begin{tabular}{c |c c c c c c c|} 
        \cline{2-8}
        & $s_{3}$ & $s_{4}$ & $s_{6}$ & $s_{1}$ & $s_{8}$ & $s_{5}$ & $s_{2}$ \\
        \hline
        \multicolumn{1}{|l|}{$s_{3}$} & \cellcolor{black} & 1 & 4 & 7 & 11 & 14 & \textbf{19} \\ 
        \multicolumn{1}{|l|}{$s_{4}$} & & \cellcolor{black} & 2 & 5 & 9 & 12 & \textbf{17} \\ 
        \multicolumn{1}{|l|}{$s_{6}$} & & & \cellcolor{black} & 2 & 6 & 9 & \textbf{14} \\ 
        \multicolumn{1}{|l|}{$s_{1}$} & & & & \cellcolor{black} & 3 & 6 & \textbf{11} \\ 
        \multicolumn{1}{|l|}{$s_{8}$} & & & & & \cellcolor{black} & 2 & \textbf{7} \\ 
        \multicolumn{1}{|l|}{$s_{5}$} & & & & & & \cellcolor{black} & \textbf{4} \\
        \multicolumn{1}{|l|}{$s_{2}$} & & & & & & & \cellcolor{black} \\ 
        \hline
    \end{tabular}
    \end{center}
    \end{table}
    Since the numbers are decreasing, the judgments are consistent and, by applying the consistency condition presented in \Cref{eq:2}, the remainder values in the table can be filled by transitivity. It is important to note, however, that the numbers introduced in the last column of \Cref{table:weights} represent a qualitative judgment regarding the \textit{closeness} between the reference criterion and the remainder, and not a quantitative judgment.
    \item Then, the DM has been asked to establish a relation between $s_2 = (0,100,0,0,0,0,0)$ (the first in the ranking) and $s_3 = (0,0,100,0,0,0,0)$ (the last). This is called the $z-$ratio value. There are several ways to obtain the value of $z$. In this study, the indifference relationship between the two reference criteria, $s_2$ and $s_3$, has been used. The technique works by lowering the performance level (and respective value) of the criterion with the highest weight, until an \textit{indifference} ($\sim$) is felt by the DM between this criterion and the criterion with the lowest weight in its highest performance level. In the case of this work, the DM stated that he would be indifferent between a ship 15 years old and a ship with 5 deficiencies or fewer. Nonetheless, the DM acknowledged that this equivalence would be admissible to other values for the age of the ship, leading to the need of performing a robustness analysis to the $z$-ratio, which is presented in Section \ref{subsubsection:robustness}. From \cref{fig:g2}, $v_2(15) = 23.52$, resulting in:
    \[
        \bar{s}_2 = (0,23.52,0,0,0,0,0) \sim s_3 = (0,0,100,0,0,0,0)
    \]
    From the additive model defined in \Cref{eq:1}, we have:
    \[
        0 w_1 + 23.52 w_2 + 0 w_3 + 0 w_4 + 0 w_5 + 0 w_6 + 0 w_7 = 0 w_1 + 0 w_2 + 100 w_3 + 0 w_4 + 0 w_5 + 0 w_6 + 0 w_7 
    \]
    Thus, $23.52 w_2 = 100 w_3$, meaning that $w_2/w_3 = 100/23.52 \approx 4.25$, which is the value of $z$, also representing in this case a substitution rate.
    \item From the non-normalized weights, $w_j$, since there are two references, the highest, $w_2 = 4.25$, and the lowest, $w_3 = 1$, the value of the unit $\alpha$ can be computed. Considering \Cref{table:weights}:
    \[
        \alpha = \frac{z.w_3 - w_3}{(19+1)} = \frac{4.25 - 1}{20} = 0.1625
    \]
    \item The remaining non-normalized weights are computed as follows, again considering \Cref{table:weights}:
    \[
        \left\{
        \begin{array}{l}
            w_1 = w_3 + (7 + 1) \times \alpha = 1 + (7 + 1) \times 0.1625 = 2.3 \\
            w_4 = w_3 + (1 + 1) \times \alpha = 1 + (1 + 1) \times 0.1625 \approx 1.33 \\
            w_5 = w_3 + (14 + 1) \times \alpha = 1 + (14 + 1) \times 0.1625 \approx 3.44 \\
            w_6 = w_3 + (4 + 1) \times \alpha = 1 + (4 + 1) \times 0.1625 \approx 1.81 \\
            w_8 = w_3 + (11 + 1) \times \alpha = 1 + (11 + 1) \times 0.1625 = 2.95 \\
        \end{array}
        \right.
    \]
    \item Finally, the normalized weights, $\hat{w}_j$, are as follows:
    \[
        \left\{
        \begin{array}{l}
            \hat{w}_1 \approx 0.13 \\
            \hat{w}_2 \approx 0.25 \\
            \hat{w}_3 \approx 0.06 \\
            \hat{w}_4 \approx 0.08 \\
            \hat{w}_5 \approx 0.20 \\
            \hat{w}_6 \approx 0.11 \\
            \hat{w}_8 \approx 0.17 \\
        \end{array}
        \right.
    \]
\end{enumerate}

\subsection{Definition of categories}
\label{subsection:definitionofcategories}
\noindent As mentioned earlier in Section \ref{subsection:classificationsystem}, the classification system through which risk categories are assigned to ships is based on a hybrid model, consisting of a rule-based system, which reproduces the rules imposed by the Paris MoU SRP, and on a value-based system, whose components are presented in the previous Sections. On the one hand, the rule-based system is used to filter ships eligible to be considered low risk ships, i.e., to be assigned category $C_1$. On the other hand, the value-based system, resulting from the application of the additive model of \Cref{eq:1}, is used to differentiate high risk ships, category $C_3$, from standard risk ships, category $C_2$. An exception exists, nonetheless, regarding ships performing ``high'' in criterion $g_3$ (deficiencies), i.e., ships that did not perform at least one inspection in the last 36 months, which are automatically assigned the $C_3$ category, regardless of the value obtained through the value-based model. The established categories are presented next.

\paragraph{A. Category $C_1$ - Low Risk Ship}
A ship is eligible to be classified as a low risk ship if it fulfils the requirements established in \Cref{table:rules}.
\begin{table}[H]
\small
\centering
\caption{Rule-based system for the category $C_1$ (Low Risk Ship).}
\label{table:rules}
\begin{center}
\begin{tabular}{|c |c c c c c c c c c|} 
    \hline 
    \multirow{1}{*}{Criterion} & \multirow{1}{*}{$g_1$} & \multirow{1}{*}{$g_2$} & \multirow{1}{*}{$g_3$} & \multirow{1}{*}{$g_4$} & \multirow{1}{*}{$g_5$} & \multirow{1}{*}{$g_6$} & \multirow{1}{*}{$g_7$} & \multirow{1}{*}{$g_8$} & \multirow{1}{*}{$g_9$} \\
    \hline
    \multirow{1}{*}{Performance level} & \multirow{1}{*}{any} & \multirow{1}{*}{any} & \multirow{1}{*}{low} & \multirow{1}{*}{no} & \multirow{1}{*}{high} & \multirow{1}{*}{high} & \multirow{1}{*}{yes} & \multirow{1}{*}{high} & \multirow{1}{*}{yes} \\
    \hline
\end{tabular}
\end{center}
\end{table}

The rules presented in \Cref{table:rules} establish that ships of any type and age are eligible to be assigned to category $C_1$, criteria $g_1$ and $g_2$, respectively; have to have less than or at most 5 deficiencies recorded in at least one inspection in the last 36 months, criterion $g_3$; must not have any detention in the last 36 months, criterion $g_4$; have to be registered in a high performance company, criterion $g_5$; have to be registered in a White flag State with IMO audit, criteria $g_6$ and $g_7$, respectively; and, finally, have to be certified by a high performance RO, recognised by at least one member State, criteria $g_8$ and $g_9$, respectively. 

\paragraph{B. Category $C_2$ - Standard Risk Ship}
A ship is considered to be a standard risk ship if, through the application of the additive model presented in \Cref{eq:1}, its performance in the different criteria adds up to a value higher than 40, i.e., $v(a) > 40$. The value of 40, serving as a cutoff value, $\lambda$, has been subjectively defined by the DM involved in the study, nonetheless, a robustness analysis regarding this value is presented in Section \ref{subsubsection:robustness}. 

\paragraph{C. Category $C_3$ - High Risk Ship}
A ship is considered to be a high risk ship if, through the application of the additive model presented in \Cref{eq:1}, its performance in the different criteria adds up to a value less than or equal to 40, i.e., $v(a) \leqslant 40$. In addition, a ship which has not performed a PSC inspection in the last 36 months (level ``high'' on criterion $g_3$), is automatically assigned to category $C_3$.

\subsection{Results and discussion}
\label{subsection:resultsanddiscussion}
\noindent The following results have been obtained through the application of the proposed MCDA approach to the data sample presented in Section \ref{subsection:definitionofperformances}. In particular, \Cref{table:resultsvalues} presents the values obtained through the application of the additive model formulated in \Cref{eq:1} to the performances of each ship presented in \Cref{table:performances}.

\begin{table}[H]
\footnotesize
\def\arraystretch{1.5}
\newcolumntype{L}[1]{>{\raggedright\let\newline\\\arraybackslash\hspace{0pt}}m{#1}}
\newcolumntype{C}[1]{>{\centering\let\newline\\\arraybackslash\hspace{0pt}}m{#1}}
\centering
\caption{Values for the ten ships in the collected data sample.}
\label{table:resultsvalues}
\resizebox{\textwidth}{!}{
\begin{tabularx}{1\textwidth} { 
    | >{\centering\arraybackslash}X
    | >{\centering\arraybackslash}X |
     >{\centering\arraybackslash}X
     >{\centering\arraybackslash}X
     >{\centering\arraybackslash}X
     >{\centering\arraybackslash}X
     >{\centering\arraybackslash}X
     >{\centering\arraybackslash}X
     >{\centering\arraybackslash}X
     >{\centering\arraybackslash}X
     >{\centering\arraybackslash}X |
     >{\raggedleft\arraybackslash}X | }
    \hline
    \multicolumn{1}{|c|}{Category} & \multicolumn{1}{c|}{Ship} & \multicolumn{1}{c}{$g_{1}$} & \multicolumn{1}{c}{$g_{2}$} & \multicolumn{1}{c}{$g_{3}$} & \multicolumn{1}{c}{$g_{4}$} & \multicolumn{1}{c}{$g_{5}$} & \multicolumn{1}{c}{$g_{6}$} & \multicolumn{1}{c}{$g_{7}$} & \multicolumn{1}{c}{$g_{8}$} & \multicolumn{1}{c|}{$g_{9}$} & \multicolumn{1}{c|}{Total} \\
    \hline
    $C_1$ & $a_{4}$ & 0.00 & 21.96 & 5.86 & 7.76 & 20.13 & 10.61 & yes & 17.28 & yes & 83.60 \\
    $C_1$ & $a_{10}$ & 0.00 & 10.54 & 5.86 & 7.76 & 20.13 & 10.61 & yes & 17.28 & yes & 72.18 \\
    \hline
    $C_2$ & $a_{1}$ & 13.47 & 3.22 & 5.86 & 7.76 & 7.55 & 10.61 & yes & 17.28 & yes & 65.75 \\
    $C_2$ & $a_{2}$ & 13.47 & 4.10 & 5.86 & 7.76 & 7.55 & 10.61 & yes & 17.28 & yes & 66.63 \\
    $C_2$ & $a_{3}$ & 13.47 & 15.22 & 2.20 & 7.76 & 7.55 & 10.61 & yes & 17.28 & yes & 74.09 \\
    $C_2$ & $a_{5}$ & 13.47 & 11.71 & 5.86 & 7.76 & 7.55 & 10.61 & yes & 17.28 & yes & 74.24 \\
    $C_2$ & $a_{7}$ & 0.00 & 10.54 & 5.86 & 7.76 & 7.55 & 10.61 & yes & 17.28 & yes & 59.59 \\
    $C_2$ & $a_{8}$ & 0.00 & 5.85 & 2.20 & 7.76 & 7.55 & 10.61 & yes & 17.28 & yes & 51.25 \\
    $C_2$ & $a_{9}$ & 13.47 & 0.00 & 5.86 & 7.76 & 7.55 & 10.61 & no & 17.28 & yes & 62.53 \\
    \hline
    $C_3$ & $a_{6}$ & 0.00 & 0.88 & 2.20 & 7.76 & 0.00 & 10.61 & yes & 17.28 & yes & 38.73 \\
    \hline
\end{tabularx}
}
\end{table}

Ships $a_4$ and $a_{10}$ are assigned to the $C_1$ category, thus, being considered low risk ships. Comparing the performances of \Cref{table:performances} with the rules in \Cref{table:rules}, it can be observed that these are the only ships fulfilling the requirements presented in the latter. On the opposite side, ship $a_6$ is the only ship assigned to the $C_3$ category, being considered a high risk ship. This is due to the total value obtained by the ship on the additive model being less than or equal to 40 (38.73). Finally, all the remainder ships are assigned to the $C_2$ category, being considered standard risk ships. 

An interesting observation is worth noting. Although ships $a_3$ and $a_5$ obtained a higher total value than ship $a_{10}$, the former are assigned to the $C_2$ category (standard risk ship), while the latter is assigned to $C_1$ (low risk ship). Taking into account \Cref{table:performances} and \Cref{table:rules}, ship $a_3$ does not fulfil the requirements regarding the criteria $g_3$ (deficiencies) and $g_5$ (company performance), and ship $a_5$ only misses the fulfilment of criterion $g_5$. Considering these results, instead of a strict rule-based system for the assignment of category $C_1$, as the model established by the Paris MoU SRP, the value-based system developed in this study could be combined with some of the established rules, for the establishment of a true hybrid model. In other words, a cutoff value for the total value obtained from the application of the additive model presented in Section \ref{subsection:aggregationmodel} to the criteria with associated value functions, i.e., all criteria except $g_7$ (IMO audit) and $g_9$ (RO recognised), could be established to differentiate category $C_1$ from category $C_2$, in addition to the rules imposed on criteria $g_7$ and $g_9$. For example, if a cutoff value were to be established between $C_1$ and $C_2$, $\lambda_{12}$, with a value of 70, for instance, ships with total values greater than 70, i.e., $v(x) > \lambda_{12} = 70$, and fulfilling the criteria $g_7$ and $g_9$ would be eligible to be assigned to category $C_1$. If this would be the case, ships $a_3$ and $a_5$ would be assigned to category $C_1$, together with $a_4$ and $a_{10}$.

\subsubsection{Robustness analysis}
\label{subsubsection:robustness}
\noindent The value of 40 for the cutoff between categories $C_2$ and $C_3$, $\lambda_{23}$, has been established according to the experience of the DM, as mentioned previously. The $z-$ratio value of 4.25 has also been obtained subjectively from the DM, as explained in Section \ref{subsection:weights}. In order to assess the influence of these parameters in the results, i.e., in the assignment of risk categories to ships, a robustness analysis has been performed for different values of $\lambda_{23}$ and $z$, creating, thus, several scenarios. The analysis has been performed for the ship $a_6$, since it is the only ship from the considered sample classified as $C_3$ (high risk ship) (\Cref{table:resultsvalues}) and given the proximity of its total value (38.73) to the cutoff established by the DM. The value of $\lambda_{23}$ has been changed in increments of one unit, from 35 to 45, i.e., $40-5 \leqslant \lambda_{23} \leqslant 40+5$. Then, the $z-$ratio value has been changed in increments of 0.5, from 3.25 to 5.25, i.e., $4.25-1 \leqslant z \leqslant 4.25+1$. The results are compared to the risk categories obtained with the application of the Paris MoU SRP and are presented in \Cref{table:robustness}.

\begin{table}[H]
\footnotesize
\def\arraystretch{1.5}
\newcolumntype{L}[1]{>{\raggedright\let\newline\\\arraybackslash\hspace{0pt}}m{#1}}
\newcolumntype{C}[1]{>{\centering\let\newline\\\arraybackslash\hspace{0pt}}m{#1}}
\centering
\caption{Comparison of risk categories assigned to ship $a_6$ for different values of $\lambda_{23}$ and $z$, and those assigned with the Paris MoU SRP. (Differences relatively to the Paris MoU SRP in bold.)}
\label{table:robustness}
\resizebox{\textwidth}{!}{
\begin{tabularx}{1\textwidth} { 
     >{\centering\arraybackslash}X
    | >{\centering\arraybackslash}X 
     >{\centering\arraybackslash}X
     >{\centering\arraybackslash}X
     >{\centering\arraybackslash}X
     >{\centering\arraybackslash}X
     >{\centering\arraybackslash}X
     >{\centering\arraybackslash}X
     >{\centering\arraybackslash}X
     >{\centering\arraybackslash}X
     >{\centering\arraybackslash}X
     >{\centering\arraybackslash}X |
     >{\centering\arraybackslash}X |}
    \cline{2-13}
    & \multicolumn{1}{c}{$35$} & \multicolumn{1}{c}{$36$} & \multicolumn{1}{c}{$37$} & \multicolumn{1}{c}{$38$} & \multicolumn{1}{c}{$39$} & \multicolumn{1}{c}{$40$} & \multicolumn{1}{c}{$41$} & \multicolumn{1}{c}{$42$} & \multicolumn{1}{c}{$43$} & \multicolumn{1}{c}{$44$} & \multicolumn{1}{c|}{$45$} & \multicolumn{1}{c|}{SRP} \\
    \hline
    \multicolumn{1}{|c|}{$z=3.25$} & \boldmath$C_2$ & \boldmath$C_2$ & \boldmath$C_2$ & \boldmath$C_2$ & \boldmath$C_2$ & \boldmath$C_2$ & $C_3$ & $C_3$ & $C_3$ & $C_3$ & $C_3$ & $C_3$ \\
    \multicolumn{1}{|c|}{$z=3.75$} & \boldmath$C_2$ & \boldmath$C_2$ & \boldmath$C_2$ & \boldmath$C_2$ & \boldmath$C_2$ & $C_3$ & $C_3$ & $C_3$ & $C_3$ & $C_3$ & $C_3$ & $C_3$ \\
    \multicolumn{1}{|c|}{$z=4.25$} & \boldmath$C_2$ & \boldmath$C_2$ & \boldmath$C_2$ & \boldmath$C_2$ & $C_3$ & $C_3$ & $C_3$ & $C_3$ & $C_3$ & $C_3$ & $C_3$ & $C_3$ \\
    \multicolumn{1}{|c|}{$z=4.75$} & \boldmath$C_2$ & \boldmath$C_2$ & \boldmath$C_2$ & \boldmath$C_2$ & $C_3$ & $C_3$ & $C_3$ & $C_3$ & $C_3$ & $C_3$ & $C_3$ & $C_3$ \\
    \multicolumn{1}{|c|}{$z=5.25$} & \boldmath$C_2$ & \boldmath$C_2$ & \boldmath$C_2$ & $C_3$ & $C_3$ & $C_3$ & $C_3$ & $C_3$ & $C_3$ & $C_3$ & $C_3$ & $C_3$ \\
    \hline
\end{tabularx}
}
\end{table}

From \Cref{table:robustness}, it can be observed that the classification of ship $a_6$ is relatively robust for higher values of the cutoff value $\lambda_{23}$. This is due to the total value, $v(a)$, obtained for different values of $z$. The total value of ships tend to increase with lower values of $z$, given the change in the criteria relative weights. In fact, the value of $a_6$ increases to 40.27 with $z=3.25$, from 38.73 with $z=4.25$. Hence, in \Cref{table:robustness}, a $\lambda_{23} \geqslant 41$ is needed to change the classification of ship $a_6$ from $C_2$ to $C_3$. On the contrary, for $z=5.25$, the change in the criteria relative weights results in the total value of ship $a_6$ to decrease from 38.73 to 37.66. Comparatively to $z=3.25$, a $\lambda_{23} \geqslant 38$ is enough to assign category $C_3$ to ship $a_6$.

These examples illustrate well the influence of the parameters $\lambda$ and $z$ in the application of the proposed MCDA methodology for the risk classification of ships, which can, nonetheless, be exploited by the DM to better reflect her/his subjective judgment.

\section{Case study}
\label{section:casestudy}
\noindent The proposed MCDA approach, developed as an alternative to the current Paris MoU SRP, has been applied to a data sample of PSC inspections performed in Lisbon, Portugal, a member State of the Paris MoU. Besides the presentation of the obtained results for the data sample, managerial insights are provided to discuss possible adjustments to the model and their implications to the assignment of ships to the different risk categories.

\subsection{PSC inspections in the port of Lisbon, Portugal, 2018}
\label{subsection:pscinspections}
\noindent In order to assess its applicability to PSC inspections, the proposed MCDA methodology have been applied to the complete data set of 138 inspections presented in Section \ref{subsection:definitionofperformances}.

The results presented in \Cref{table:casestudy} are obtained for the values of $\lambda_{23}$ and $z$ used previously in the robustness analysis, and serve to assess the differences relatively to the application of the current Paris MoU SRP in a real set of PSC inspections performed in a Paris MoU member State port.

\begin{table}[H]
\footnotesize
\def\arraystretch{1.5}
\newcolumntype{L}[1]{>{\raggedright\let\newline\\\arraybackslash\hspace{0pt}}m{#1}}
\newcolumntype{C}[1]{>{\centering\let\newline\\\arraybackslash\hspace{0pt}}m{#1}}
\centering
\caption{Comparison of the number of ships in each risk category for different values of $\lambda_{23}$ and $z$ and the Paris MoU SRP. (Differences relatively to the Paris MoU SRP in bold.)}
\label{table:casestudy}
\resizebox{\textwidth}{!}{
\begin{tabularx}{1\textwidth} { 
     >{\centering\arraybackslash}X
     >{\centering\arraybackslash}X 
    | >{\centering\arraybackslash}X
     >{\centering\arraybackslash}X
     >{\centering\arraybackslash}X
     >{\centering\arraybackslash}X
     >{\centering\arraybackslash}X
     >{\centering\arraybackslash}X
     >{\centering\arraybackslash}X
     >{\centering\arraybackslash}X
     >{\centering\arraybackslash}X
     >{\centering\arraybackslash}X
     >{\centering\arraybackslash}X |
     >{\centering\arraybackslash}X |}
    \cline{3-14}
    & & \multicolumn{1}{c}{$35$} & \multicolumn{1}{c}{$36$} & \multicolumn{1}{c}{$37$} & \multicolumn{1}{c}{$38$} & \multicolumn{1}{c}{$39$} & \multicolumn{1}{c}{$40$} & \multicolumn{1}{c}{$41$} & \multicolumn{1}{c}{$42$} & \multicolumn{1}{c}{$43$} & \multicolumn{1}{c}{$44$} & \multicolumn{1}{c|}{$45$} & \multicolumn{1}{c|}{SRP} \\
    \hline
    \multicolumn{1}{|c}{\multirow{3}{*}{$z=3.25$}} & \multicolumn{1}{|c|}{$C_1$} & 17 & 17 & 17 & 17 & 17 & 17 & 17 & 17 & 17 & 17 & 17 & 17 \\
    \multicolumn{1}{|c}{} & \multicolumn{1}{|c|}{$C_2$} & \textbf{119} & \textbf{119} & \textbf{119} & \textbf{119} & \textbf{119} & \textbf{119} & 118 & 118 & \textbf{117} & \textbf{117} & \textbf{116} & 118 \\
    \multicolumn{1}{|c}{} & \multicolumn{1}{|c|}{$C_3$} & \textbf{2} & \textbf{2} & \textbf{2} & \textbf{2} & \textbf{2} & \textbf{2} & 3 & 3 & \textbf{4} & \textbf{4} & \textbf{5} & 3 \\
    \hline
    \multicolumn{1}{|c}{\multirow{3}{*}{$z=3.75$}} & \multicolumn{1}{|c|}{$C_1$} & 17 & 17 & 17 & 17 & 17 & 17 & 17 & 17 & 17 & 17 & 17 & 17 \\
    \multicolumn{1}{|c}{} & \multicolumn{1}{|c|}{$C_2$} & \textbf{119} & \textbf{119} & \textbf{119} & \textbf{119} & \textbf{119} & 118 & 118 & 118 & \textbf{117} & \textbf{116} & \textbf{116} & 118 \\
    \multicolumn{1}{|c}{} & \multicolumn{1}{|c|}{$C_3$} & \textbf{2} & \textbf{2} & \textbf{2} & \textbf{2} & \textbf{2} & 3 & 3 & 3 & \textbf{4} & \textbf{5} & \textbf{5} & 3 \\
    \hline
    \multicolumn{1}{|c}{\multirow{3}{*}{$z=4.25$}} & \multicolumn{1}{|c|}{$C_1$} & 17 & 17 & 17 & 17 & 17 & 17 & 17 & 17 & 17 & 17 & 17 & 17 \\
    \multicolumn{1}{|c}{} & \multicolumn{1}{|c|}{$C_2$} & \textbf{119} & \textbf{119} & \textbf{119} & \textbf{119} & 118 & 118 & 118 & 118 & \textbf{116} & \textbf{116} & \textbf{116} & 118 \\
    \multicolumn{1}{|c}{} & \multicolumn{1}{|c|}{$C_3$} & \textbf{2} & \textbf{2} & \textbf{2} & \textbf{2} & 3 & 3 & 3 & 3 & \textbf{5} & \textbf{5} & \textbf{5} & 3 \\
    \hline
    \multicolumn{1}{|c}{\multirow{3}{*}{$z=4.75$}} & \multicolumn{1}{|c|}{$C_1$} & 17 & 17 & 17 & 17 & 17 & 17 & 17 & 17 & 17 & 17 & 17 & 17 \\
    \multicolumn{1}{|c}{} & \multicolumn{1}{|c|}{$C_2$} & \textbf{119} & \textbf{119} & \textbf{119} & \textbf{119} & 118 & 118 & 118 & \textbf{117} & \textbf{116} & \textbf{116} & \textbf{114} & 118 \\
    \multicolumn{1}{|c}{} & \multicolumn{1}{|c|}{$C_3$} & \textbf{2} & \textbf{2} & \textbf{2} & \textbf{2} & 3 & 3 & 3 & 3 & \textbf{5} & \textbf{5} & \textbf{5} & 3 \\
    \hline
    \multicolumn{1}{|c}{\multirow{3}{*}{$z=5.25$}} & \multicolumn{1}{|c|}{$C_1$} & 17 & 17 & 17 & 17 & 17 & 17 & 17 & 17 & 17 & 17 & 17 & 17 \\
    \multicolumn{1}{|c}{} & \multicolumn{1}{|c|}{$C_2$} & \textbf{119} & \textbf{119} & \textbf{119} & 118 & 118 & 118 & \textbf{117} & \textbf{116} & \textbf{116} & \textbf{116} & \textbf{114} & 118 \\
    \multicolumn{1}{|c}{} & \multicolumn{1}{|c|}{$C_3$} & \textbf{2} & \textbf{2} & \textbf{2} & 3 & 3 & 3 & \textbf{4} & \textbf{5} & \textbf{5} & \textbf{5} & \textbf{7} & 3 \\
    \hline
\end{tabularx}
}
\end{table}

From \Cref{table:casestudy}, it can be observed that, for different values of $z$, the number of ships assigned to category $C_1$ (low risk ship) in 138 inspections has been constantly 17. This is due to the fact that, like the Paris MoU SRP, the assignment of ships to the $C_1$ category is purely rule-based, independently of the value scored by each ship in the additive model (\Cref{eq:1}). On the contrary, the assignment of ships to the $C_2$ (standard risk ship) and $C_3$ (high risk ship) categories depends on the values of $\lambda_{23}$ and $z$. As shown in Section \ref{subsubsection:robustness}, the relative weights of criteria resulting from lower values of $z$ tend to increase the total value of the ships. Thus, higher values of $\lambda_{23}$ are required to change the classification of ships from $C_2$ to $C_3$. For example, with $z=3.25$, only for $\lambda_{23} \geqslant 41$ are three or more ships classified as $C_3$, while for $z=4.25$ this happens for $\lambda_{23} \geqslant 39$, and for $z=5.25$ this happens for $\lambda_{23} \geqslant 38$. In addition, for $z=3.25$, a maximum of 5 ships are classified as $C_3$, corresponding to a minimum of 116 ships classified as $C_2$. For $z=5.25$, a maximum of 7 ships classified as $C_3$ is obtained, corresponding to a minimum of 114 ships classified as $C_2$. These results further demonstrate the relevance of the parameters in the application of the proposed methodology. Nonetheless, with $39 \leqslant \lambda_{23} \leqslant 42$, for at least three out of the five values presented for $z$, the number of ships in each category obtained with the proposed model is equal to that obtained with the Paris MoU SRP, revealing some robustness for small changes in $\lambda_{23}$.

\subsection{Managerial insights}
\label{subsection:managerialinsights}
\noindent Contrarily to what is stated in the Annex 7 of the Paris MoU \citep{ParisMoU2020}, i.e., the ``Ship Risk Profile'', the criteria weights used in the SRP do not reflect ``the relative influence of each parameter on the overall risk of the ship''. Instead, they are used to measure the performance of each ship on each criterion according to its characteristics.

The proposed MCDA methodology, based on the DCM for the assignment of risk profiles to ships, represents an important departure from the current Paris MoU SRP. Through the additive model presented in \Cref{eq:1}, the methodology enables the sorting of ships according to their performances on the weighted criteria. Three important aspects regarding the application of the methodology are discussed next.

\begin{enumerate}
    \item Criteria value functions: As previously mentioned, an expert with research experience in PSC, particularly in the Paris MoU, has been consulted in this study. The application of the DCM to the SRP presented in Section \ref{section:application} reflects his subjective judgments on the matter. The criteria value functions obtained from the comparison tables (e.g. \Cref{table:g2}) represent his preference differences between the levels in the criteria scales. Different values would be obtained if other experts, acting as DMs, were to be consulted. Nonetheless, the nature of the DCM allows the subjective judgments of the DMs regarding the differences in the criteria scales to be captured and exploited, unlike the current Paris MoU SRP. In this particular case, the results of the interaction between the consulted expert and the analysts team regarding the construction of the criteria value functions are presented in Section \ref{subsection:valuefunctions} and \ref{appendix:a}, namely the computed values for the scale levels derived from the comparison tables.
    \item Criteria weights: The discussion presented before for the criteria value functions is applicable to the criteria weights. Equally important to the capability of a given method to apprehend the DM's subjectivity and to exploit it is the correctness under which this is achieved. The DCM ensures such a correctness by providing meaning to the model's components, namely the criteria weights, which represent in this case trade-offs between criteria, or, in other words, substitution rates. The value of $z$, the ratio between two reference criteria, in this case the first and the last ranked criteria, $s_2$ and $s_3$, respectively, can be obtained through different approaches. In this study, the indifference between the reference criteria has been used, an innovative aspect of the application of the DCM. The $z$-ratio decisively affects the definition of the criteria weights, thus, being important to be correctly established. The results of the interaction between the consulted expert and the analysts team regarding the definition of the criteria weights are presented in Section \ref{subsection:weights}. The expert considered that a ship with the highest performance on criterion $g_2$ (age of ship) would be strictly preferred to any ship with the highest performance on any other criteria. A ship with the highest performance on criterion $g_3$ (deficiencies) was considered the least preferred. By exploiting the indifference felt by the expert between both alternatives, a $z$-ratio of 4.25 have been obtained, through which the criteria weights have been computed.
    \item Definition of categories: Regarding the categories, the same risk categories as those defined in the Paris MoU have been used in this study. In addition, category $C_1$ (low risk ship), has been defined according to a rule-based system similar to that established in the Paris MoU SRP. Categories $C_2$ (standard risk ship) and $C_3$ (high risk ship) are assigned according to the scores obtained by the ships in the additive model of \Cref{eq:1}. If a ship obtains a score less than or equal to a given cutoff value, $\lambda_{23}$, the ship is assigned to category $C_3$; if the score is greater than the same value, the ship is assigned to category $C_2$. First, instead of a rule-based system for the definition of category $C_1$, the defined additive model could be used in combination with acceptation/rejection criteria (for criteria $g_7$ and $g_9$ in particular), if a cutoff value were to be established between category $C_1$ and category $C_2$, $\lambda_{12}$. Second, the used value for $\lambda_{23}$, equal to 40, has been, again, subjectively chosen by the DM. Similarly to the aforementioned parameters, this value can be adjusted to reflect the subjective judgment of different DMs. As for the considered risk categories, presented in Section \ref{subsection:definitionofcategories}, the consulted expert provided a cutoff value of 40 between categories $C_2$ and $C_3$. Considering $z=4.25$, a value of 40 for $\lambda_{23}$ resulted exactly in the same risk classification as that obtained through the Paris MoU SRP, using a data sample of 138 PSC inspections at a Paris MoU member State port (\Cref{table:casestudy}).
\end{enumerate}

The aspects discussed above should be taken into account when applying the proposed methodology for the assignment of risk profiles to ships within the scope of PSC inspections. The chosen values for the parameters will influence the classification of ships, as demonstrated by the robustness analysis performed in Section \ref{subsubsection:robustness}. This fact can be exploited according to the DM's interest. For example, if a stricter SRP is to be imposed, higher values for the $z$-ratio and for the categories' cutoff should be chosen. On the contrary, if a more permissive SRP is to be implemented, the opposite should occur. However, these decisions should be thoughtfully considered, since the risk classification of ships determines the frequency and level of detail of PSC inspections, which will require the availability of resources from PSC authorities.

\section{Conclusions and future research}
\label{section:conclusion}
\noindent The elimination of substandard shipping is an objective of maritime authorities worldwide, given the social, economic, and environmental consequences that may result from the non compliance with international safety rules at sea. In European waters, the enforcement of international conventions on maritime safety is ensured by the Paris MoU, the first regional agreement on PSC. In particular, the selection scheme for PSC inspection is determined by the Paris MoU SRP, under which ships are assigned points according to their performances in a set of criteria. Depending on the obtained score, or on the fulfilment of a set of rules, each ship is classified with a ``risk profile''. Being in essence an MCDA tool to solve a sorting problem, comprising a set of criteria, criteria weights, and a set of categories to which actions are assigned, limited research has been performed on the Paris MoU SRP from an MCDA perspective. In addition, the SRP criteria weights do not present a clear meaning, nor origin. This study uses the DCM as an alternative to the SRP for the risk classification of ships. Instead of being used solely for the establishment of criteria weights as it was originally conceived, the DCM is used to build a complete MCDA model, an innovation relatively to the reviewed literature and a contribution of this study.

Counting with the participation of an expert on maritime policy, and using a data set of 138 PSC inspections performed in the port of Lisbon, Portugal, in 2018, the DCM is used in the construction of an MCDA model for ship risk classification on the following terms. First, the criteria established in the Paris MoU SRP are organised into a framework comprising \textit{points of view}, \textit{significance axes}, and \textit{criteria} with associated \textit{scales}. Second, the DCM is used to define value functions for the considered criteria, both for those with discrete scales and that with a continuous scale. Third, criteria weights are established, again, through the DCM. A novel aspect of the study is that the $z$-ratio, a parameter that allows the determination of all criteria weights, instead of being directly elicited from the DM, is obtained by exploring the indifference of the DM between two reference criteria. Finally, based on the ship risk profiles established in the Paris MoU SRP, risk categories are defined. Similarly to the SRP, the low risk category, category $C_1$, is assigned through a rule-based system, while the remainder two, the standard and the high risk categories, categories $C_2$ and $C_3$, respectively, are differentiated through the definition of a cutoff value, $\lambda$. Results are then obtained and discussed. In addition, a robustness analysis is performed on the variation of the model's parameters, namely the $z$-ratio and the cutoff value $\lambda$.

Relevant conclusions can be drawn from the application of the proposed methodology to the collected data sample, as well as managerial insights can be provided. The purpose of using the DCM in this study was to allow the subjective judgment of the DM to be captured and exploited in the construction of the MCDA model. The elements presented in this paper, from the criteria framework to the criteria scales, from the criteria value functions to the criteria weights, reflect the subjective judgment of the consulted expert. These can be adjusted through the DCM to reflect different judgments. In particular, the values of $z$ and $\lambda$ can be adjusted in order to implement a stricter or, otherwise, more permissive ship risk classification system. However, attention should be given to these values as the resulting classification will demand more or less resources from PSC authorities.

As for future research, multiple paths can be followed. First, regarding the application of the DCM as an alternative to the Paris MoU SRP, the following can be considered: criteria have been assumed as independent in the present study, but that may not be the case and interactions between criteria should be explored; this study presents a novel elicitation technique for the $z$-ratio, which is used in the definition of the criteria weights, but the DecSpace platform\footnote{Available at \url{http://decspace.sysresearch.org/index.html}} should be tested for this purpose; robustness analyses can be performed through simulation in order to cover more scenarios for different model's parameters; and the proposed model should be validated with different DMs. Second, alternative MCDA methods should be applied to the ship risk classification problem, such as the \textsc{Electre}  outranking method. This should be done in order to assess and compare the obtained results and the effort spent in the model implementation.

\section*{Acknowledgements}
\noindent José Rui Figueira gratefully acknowledge the financial support of Fundação para a Ciência e a Tecnologia under the WISDom research project (grant number DSAIPA/DS/0089/2018), through the Data Science and Artificial Intelligence in Public Administration Programme. The contribution of the third author was co-funded by the European Regional Development Fund (Fundo Europeu de Desenvolvimento Regional - FEDER) and by the Portuguese Foundation for Science and Technology (Fundação para a Ciência e a Tecnologia - FCT) under project ``Integrated System for Traffic Monitoring and Maritime Risk Assessment (MoniRisk)'', No. 028746.


\cleardoublepage

\appendix
\section{Comparison tables and value functions for the considered criteria.}
\label{appendix:a}
\begin{enumerate}
    \item Criterion $g_{1}$ ACCI (Ship accident consequences). Developed in the article.
    \item Criterion $g_{2}$ AGES (Age of ship). Developed in the article.
    \item Criterion $g_{3}$ DEFC (Deficiencies). The number of blank cards added by the DM between the levels of criterion $g_{3}$ is presented in \Cref{table:g3} in bold, which respect the consistency condition established in \Cref{eq:2}.
    \begin{table}[H]
        \small
        \centering
        \caption{Criterion $g_3$ (deficiencies) comparison table.}
        \label{table:g3}
        \begin{center}
            \begin{tabular}{c |c c c|} 
                \cline{2-4}
                & $l_{3,3}$ & $l_{3,2}$ & $l_{3,1}$ \\
                \hline
                \multicolumn{1}{|l|}{$l_{3,3}$} & \cellcolor{black} & \textbf{2} & 7 \\ 
                \multicolumn{1}{|l|}{$l_{3,2}$} & & \cellcolor{black} & \textbf{4} \\ 
                \multicolumn{1}{|l|}{$l_{3,1}$} & & & \cellcolor{black} \\ 
                \hline
            \end{tabular}
        \end{center}
    \end{table}
    Considering $v_3(l_{3,1})=100$ and $v_3(l_{3,3})=0$ results in:
    \[
        \alpha = \frac{v_3(l_{3,1}) - v_3(l_{3,3})}{(2+1)+(4+1)} = \frac{100 - 0}{8} = 12.5
    \]
    and, consequently:
    \[
        v_3(l_{3,2}) = v_3(l_{3,3}) + (2 + 1) \times \alpha = 0 + (2 + 1) \times 12.5 = 37.5
    \]
    The scale for criterion $g_3$ is presented in \Cref{criteriascales:b}.
    
    \item Criterion $g_{4}$ DETN (Detentions). The preference differences established by the DM for the scale levels of criterion $g_{4}$ are presented in bold in \Cref{table:g4}, which respect the consistency condition established in \Cref{eq:2}.
    \begin{table}[H]
        \small
        \centering
        \caption{Criterion $g_4$ (detentions) comparison table.}
        \label{table:g4}
        \begin{center}
            \begin{tabular}{c |c c c|} 
                \cline{2-4}
                & $l_{4,3}$ & $l_{4,2}$ & $l_{4,1}$ \\
                \hline
                \multicolumn{1}{|l|}{$l_{4,3}$} & \cellcolor{black} & \textbf{3} & 8 \\ 
                \multicolumn{1}{|l|}{$l_{4,2}$} & & \cellcolor{black} & \textbf{4} \\ 
                \multicolumn{1}{|l|}{$l_{4,1}$} & & & \cellcolor{black} \\ 
                \hline
            \end{tabular}
        \end{center}
    \end{table}
    Considering $v_4(l_{4,1})=100$ and $v_4(l_{4,3})=0$ results in:
    \[
        \alpha = \frac{v_4(l_{4,1}) - v_4(l_{4,3})}{(3+1)+(4+1)} = \frac{100 - 0}{9} \approx 11.11
    \]
    and, consequently:
    \[
        v_4(l_{4,2}) = v_3(l_{3,3}) + (3 + 1) \times \alpha = 0 + (3+ 1) \times 11.11 \approx 44.44
    \]
    The scale for criterion $g_4$ is presented in \Cref{criteriascales:c}.
    
    \item Criterion $g_{5}$ COPF (Company performance). The number of blank cards added by the DM between the levels of criterion $g_{5}$ is presented in \Cref{table:g5} in bold, which are the same as in criterion $g_{3}$. As such, $\alpha = 12.5$ and $v_3(l_{3,2}) = 37.5$. The scale for criterion $g_5$ is presented in \Cref{criteriascales:d}.
    \begin{table}[H]
        \small
        \centering
        \caption{Criterion $g_5$ (company performance) comparison table.}
        \label{table:g5}
        \begin{center}
            \begin{tabular}{c |c c c|} 
                \cline{2-4}
                & $l_{5,1}$ & $l_{5,2}$ & $l_{5,3}$ \\
                \hline
                \multicolumn{1}{|l|}{$l_{5,1}$} & \cellcolor{black} & \textbf{2} & 7 \\ 
                \multicolumn{1}{|l|}{$l_{5,2}$} & & \cellcolor{black} & \textbf{4} \\ 
                \multicolumn{1}{|l|}{$l_{5,3}$} & & & \cellcolor{black} \\ 
                \hline
            \end{tabular}
        \end{center}
    \end{table}    
    \item Criterion $g_{6}$ FLPF (Flag performance). The preference differences established by the DM for the scale levels of criterion $g_{6}$ are presented in bold in \Cref{table:g6}, which respect the consistency condition established in \Cref{eq:2}.
    \begin{table}[H]
        \small
        \centering
        \caption{Criterion $g_6$ (flag performance) comparison table.}
        \label{table:g6}
        \begin{center}
            \begin{tabular}{c |c c c c|} 
                \cline{2-5}
                & $l_{6,1}$ & $l_{6,2}$ & $l_{6,3}$ & $l_{6,4}$ \\
                \hline
                \multicolumn{1}{|l|}{$l_{6,1}$} & \cellcolor{black} & \textbf{2} & 7 & 14 \\ 
                \multicolumn{1}{|l|}{$l_{6,2}$} & & \cellcolor{black} & \textbf{4} & 11 \\ 
                \multicolumn{1}{|l|}{$l_{6,3}$} & & & \cellcolor{black} & \textbf{6} \\
                \multicolumn{1}{|l|}{$l_{6,4}$} & & & & \cellcolor{black} \\ 
                \hline
            \end{tabular}
        \end{center}
    \end{table}
    Considering $v_6(l_{6,4})=100$ and $v_6(l_{6,1})=0$ results in:
    \[
        \alpha = \frac{v_6(l_{6,4}) - v_6(l_{6,1})}{(2+1)+(4+1)+(6+1)} = \frac{100 - 0}{15} \approx 6.6667
    \]
    and, consequently:
    \begin{itemize}
        \item[] $v_6(l_{6,2}) = v_6(l_{6,1}) + (2 + 1) \times \alpha = 0 + (2 + 1) \times 6.6667 \approx 20.00$
        \item[] $v_6(l_{6,3}) = v_6(l_{6,1}) + (7 + 1) \times \alpha = 0 + (7 + 1) \times 6.6667 \approx 53.33$
    \end{itemize}
    The scale for criterion $g_6$ is presented in \Cref{criteriascales:e}.
    
    \item Criterion $g_{7}$ FLIA (Fulfilment of the IMO Audit). Criterion $g_{7}$ is an acceptance/rejection criterion, thus, not having an associated comparison table, nor value function. 
    
    \item Criterion $g_{8}$ ROPF (RO performance). The preference differences established by the DM for the scale levels of criterion $g_{8}$ are presented in \Cref{table:g8} in bold, which also respect the consistency condition established in \Cref{eq:2}.
    \begin{table}[H]
        \small
        \centering
        \caption{Criterion $g_8$ (RO performance) comparison table.}
        \label{table:g8}
        \begin{center}
            \begin{tabular}{c |c c c|} 
                \cline{2-4}
                & $l_{8,1}$ & $l_{8,2}$ & $l_{8,3}$ \\
                \hline
                \multicolumn{1}{|l|}{$l_{8,1}$} & \cellcolor{black} & \textbf{3} & 7 \\ 
                \multicolumn{1}{|l|}{$l_{8,2}$} & & \cellcolor{black} & \textbf{3} \\ 
                \multicolumn{1}{|l|}{$l_{8,3}$} & & & \cellcolor{black} \\ 
                \hline
            \end{tabular}
        \end{center}
    \end{table}
    Considering $v_8(l_{8,3})=100$ and $v_8(l_{8,1})=0$ results in:
    \[
        \alpha = \frac{v_8(l_{8,3}) - v_8(l_{8,1})}{(3+1)+(3+1)} = \frac{100 - 0}{8} = 12.5
    \]
    and, consequently:
    \[
        v_8(l_{8,2}) = v_8(l_{8,1}) + (3 + 1) \times \alpha = 0 + (3+ 1) \times 12.5 = 50
    \]
    The scale for criterion $g_8$ is presented in \Cref{criteriascales:f}.
    
    \item Criterion $g_{9}$ AGES (RO recognised by at least one member state). Like criterion $g_{7}$, criterion $g_{9}$ is an acceptance/rejection criterion, not having an associated comparison table, nor value function.
    
\end{enumerate}

\end{document}